\documentclass[showpacs,preprintnumbers,amsmath,amssymb]{revtex4}
\usepackage[dvips]{graphicx}
\usepackage{dcolumn}

\begin{document}

\title
{Anisotropic Strong Coupling Effects on Superfluid $^3$He in Aerogels
 
 \ Conventional Spin-Fluctuation Approach \ } 

\author{Ryusuke Ikeda}

\affiliation{%
Department of Physics, Kyoto University, Kyoto 606-8502, Japan
}

\date{\today}

\begin{abstract} 
Motivated by recent experiments on liquid $^3$He reporting emergence of novel superfluid phases in globally anisotropic aerogels, our previous theory on superfluid $^3$He in globally anisotropic aerogels is extended to incorporate effects of an anisotropy of the quasiparticle scattering cross section on the strong coupling (SC) contributions to the Ginzburg-Landau (GL) free energy on the basis of the spin-fluctuation (paramagnon) approach to the SC contributions developed by Brinkman et al. [Phys. Rev. A {\bf 10}, 2386 (1974)]. In the globally isotropic case, impurity effects on the SC correction destabilize the A-phase even at higher pressures of about 30 (bar) and make the B-phase the only state in equilibrium, while SC contributions accompanied by a global stretched anisotropy to the GL quartic terms generally tend to broaden the stability region of the A-phase compared with that of the B-phase. In particular, in contrast to the cases in bulk and in the isotropic aerogel, the SC corrections to the GL {\it quadratic} terms are not negligible in the globally anisotropic case but may change the sign of the apparent anisotropy depending on the magnitude of the frequency cutoff of the normal paramagnon propagator. Based on this sign change of the apparent anisotropy, we discuss different strange observations on superfluid $^3$He in porous media such as disappearance of the polar superfluid phase at higher pressures seen in nematically-ordered aerogels and absence of the B-phase and of the A-phase with planar ${\hat {\bf l}}$-vector in a stretched aerogel. 
\end{abstract}



\maketitle

\section{Introduction}
Anisotropy added to Fermi superfluids with isotropic Fermi surface has profound effects on realization of Cooper-pairing states. The isotropic superfluid $^3$He of the bulk liquid has been thoroughly studied so far \cite{VW} and was shown to, in zero field, consist of just two pairing states, Anderson-Brinkman-Morel (ABM) state \cite{ABM,BSA} and Balian Werthamer (BW) one \cite{BW}. It has been suggested \cite{HalperinJPSJ,Saulsv1,Yoonseok} that superfluid $^3$He in an aerogel suffers from scattering events due to the surface of the {\it locally} two-dimensional (2D) porous material and will enhance stability of the ABM state, i.e., the A-phase. The ensuing theoretical study on effects of a {\it global} anisotropy of scattering events on superfluid $^3$He based on the impurity-scattering model \cite{Saulsv1} has shown an enhanced stability of A-phase and, more importantly, realization of the polar pairing state in the so-called stretched (or, 1D-like) aerogel where the quasiparticle mean-free path is longer along the uniaxial anisotropy axis \cite{AI06}. Reflecting the fact that the polar pairing state gains a strong coupling (SC) contribution of the same extent as the ABM state in the isotropic bulk liquid \cite{VWcom}, this polar phase just below the superfluid transition temperature $T_c(P)$ does not shrink with increasing $P$, where $P$ is the pressure. 

Recent experiments on superfluid $^3$He in aerogels have suggested the presence of novel pairing states essentially differing from those appearing in the bulk liquid when the aerogel has a well-defined global anisotropy \cite{DmitrievLT,HalperinNP,HalperinPRL}. First, as convincingly argued in Ref.\cite{AI06}, the polar pairing state has been realized in 1D-like nematically-ordered aerogels with lower porosities \cite{DmitrievLT}. However, the resulting phase diagrams on the nematically-ordered aerogels have shown the polar state disappearing with increasing $P$, in contrast to the tendency following from the conventional SC corrections \cite{AI06,VWcom} mentioned above. Further, an experiment in a different stretched (1D-like) aerogel has shown a surprising phase diagram: There, both the polar and BW pairing states are absent, and just two equal spin pairing (ESP) states are found, the familiar A-phase at higher temperatures and a biaxial A-like phase at lower temperatures \cite{Sauls}. In addition, in the corresponding compressed (2D-like) aerogel, the A-phase has been found only in higher magnetic fields than a threshold field. The absence of the A-phase in zero field in Ref.\cite{HalperinPRL} implies that the naive picture \cite{HalperinJPSJ,Saulsv1,Yoonseok} based on the locally 2D-like nature does not hold. Another unexpected fact in these two experiments \cite{HalperinNP,HalperinPRL} is that the ${\hat {\bf l}}$-vector in the high temperature A-phases has counterintuitive orientations: Contrary to the conventional picture \cite{AI06,Volovik08,Sauls}, the ${\hat {\bf l}}$-vector in the 1D-like aerogel is parallel to the uniaxial anisotropy axis, while it lies in the perpendicular plane to the axis in the 2D-like compressed aerogel. Clearly, some factor is lacking for describing superfluid $^3$He in anisotropic aerogels even in the picture based on the conventional impurity scattering model \cite{Saulsv1,AI06}. 

In this work, theory of superfluid $^3$He in globally anisotropic aerogel is extended by incorporating effects of the anisotropic scattering of quasiparticles in the {\it SC contribution} to the free energy. The global anisotropy is measured by a parameter $\delta_u$, and the case with a positive (negative) $\delta_u$ describes a compressed (stretched) aerogel. Throughout this paper, we restrict ourselves to the use of the conventional spin-fluctuation (paramagnon) model \cite{BSA} in describing the SC contributions to the Ginzburg-Landau (GL) free energy, because the dynamics (frequency dependence) of the effective interaction between the normal quasiparticles becomes important. In the previous theory \cite{AI06} on the anisotropic case, effects of the global anisotropy on the SC correction have been simply assumed to be negligibly small. Here, we show that, in the $\delta_u \neq 0$ case with a global anisotropy, the SC correction to the {\it quadratic} term in the GL free energy determining the mean field superfluid transition temperature $T_c(\delta_u)$ is not negligible but plays important roles. In particular, this SC correction to the quadratic term has the {\it opposite} pressure dependence to the corresponding weak-coupling term derived in \cite{AI06}. We will argue that the resulting anisotropy effect on $T_c$ is the origin of various unexpected observations in stretched or 1D-like aerogels, i.e., the observed $P$-dependence of the polar phase region in Ref.\cite{DmitrievLT}, the absence of the polar phase \cite{HalperinNP}, and the counterintuitive orientations of the ${\hat {\bf l}}$-vector \cite{HalperinNP,HalperinPRL}. 

Next, to construct the theoretical phase diagrams fully, the SC contributions to the quartic terms of the GL free energy are also examined. In general, the quartic SC correction is present at the lowest order of the anisotropy parameter $\delta_u$ so that stability of the ABM state is significantly affected by the sign of $\delta_u$ for a large anisotropy. We find that the SC contributions to the coefficients of the GL quartic terms tend to increase with increasing $\delta_u$, implying that the ABM state tends to be stabilized (destabilized) in uniaxially stretched (compressed) aerogels. This obtained feature is qualitatively consistent with that seen in experiments of the Northwestern University group \cite{HalperinNP,HalperinPRL}. 

To compare theoretical results with the experimental observations \cite{DmitrievLT,HalperinNP,HalperinPRL}, we have examined parameter dependences of the resulting phase diagrams. As parameters affecting the phase diagram, there are three parameters to be changed in the present approach, which are the anisotropy, the averaged mean-free path or the disorder strength, and an upper cutoff 
$E_c$ for the frequency carried by the paramagnon propagator. This cutoff $E_c$ inevitably appears in the present approach where the conventional treatment in Ref.\cite{BSA} for the bulk liquid is applied to the case in aerogels. The resulting phase diagram unexpectedly depends on the magnitude of $E_c$, because the magnitude of $E_c$ determines the pressure value above which the superfluid phases with the above-mentioned counterintuitive anisotropy and orientation of the ${\hat {\bf l}}$-vector occur. Broadly speaking, for a large enough $E_c$, we obtain phase diagrams in the stretched aerogel case with no polar and BW states which are qualitatively consisitent 
with that in Ref.\cite{HalperinNP}, while the phase diagram following from an $E_c$ of a moderate magnitude is similar to that in the previous work \cite{AI06} but with the polar phase shrinking with increasing $P$ and thus, is consistent with the observed ones in Ref.\cite{DmitrievLT}. At present, it is unclear what this $E_c$ dependence of the resulting phase diagram implies, and  a further study based on other models of the SC contribution would be needed. 

This paper is organized as follows. In $\S 2$, the model hamiltonian and the previous results in Ref.\cite{AI06} where no anisotropic SC contributions were considered are reviewed. In $\S 3$, the conventional paramagnon approach to the SC contributions is reviewed and is applied to the evaluation of a large isotropic SC contribution which convincingly results in the absence of the A phase in equilibrium in the globally isotropic aerogel. In $\S 4$, the SC contributions to the GL free energy terms in the globally anisotropic cases are calculated. In $\S 5$, the resulting phase diagrams are discussed in details, and a summary and comments are given in the last section. Details of the obtained coefficients are given in Appendix.

\section{Model and Review on Previous Results}

In this section, we introduce the microscopic models and review results on the GL free energy in the weak-coupling approximation \cite{AI06}. The SC correction will be considered in the following sections. 

We use the BCS Hamiltonian including the additional term 
\begin{equation}
{\cal H}_{\rm imp} = \int d^3{\bf r} \sum_\sigma {\hat \psi}^\dagger_\sigma({\bf r}) \, u({\bf r}) \, {\hat \psi}_\sigma({\bf r}), 
\end{equation} 
associated with the impurity scattering with a potential $u({\bf r})$, where ${\hat \psi}$ is the fermion operator. In the cases with a global uniaxial anisotropy with the anisotropy axis along ${\hat z}$, the random potential $u({\bf r})$ is assumed to have zero mean and to satisfy 
\begin{equation}
{\overline {|u_{\bf k}|^2}} = \frac{1}{2 \pi N(0) \tau_o}(1 + \delta_u {\hat {\bf k}}_z^2)
\label{impline}
\end{equation}
in the Fourier representation, where the overbar denotes the random average, $N(0)$ is the density of states per spin on the Fermi surface in the normal state, and ${\hat {\bf k}}={\bf k}/k_{\rm F}$ with the Fermi wavenumber $k_{\rm F}$. The random-averaged Matsubara Green's function defined in the normal state takes the form 
\begin{equation}
{\cal G}_\varepsilon({\bf p}) = \frac{1}{{\rm i}\varepsilon - \xi_{\bf p} + {\rm i} {\rm sgn}(\varepsilon) \eta_{\bf p}}, 
\end{equation}
where $\varepsilon$ is a fermionic Matsubara frequency, and 
\begin{equation}
\eta_{\bf p} = \frac{1}{2 \tau_o}(1 + \delta_u {\hat p}_z^2). 
\label{damping}
\end{equation}
Based on this, $\tau_o$ appeared in eqs.(\ref{impline}) and (\ref{damping}) is regarded as the relaxation time of a single quasiparticle in the isotropic case. 
Then, the difference in the free energy density between the superfluid and normal states is written as 
\begin{equation}
F = \biggl\langle \frac{1}{|g|} {\rm Tr} (\Delta^\dagger({\hat p}) \Delta({\hat p}) ) \biggr\rangle_{\hat p} - T
{\rm ln} \langle T_s \exp{\Pi} \rangle, 
\label{bcsint}
\end{equation}
where 
\begin{equation}
\Pi = \sum_{\bf p} 
\biggl[ (\Delta^\dagger ({\hat p}) )_{\beta \alpha} 
\int_0^{1/T} ds \, a_{p, \alpha}(s) \, 
a_{-p, \beta}(s) 
\biggr] + {\rm h.c.}, 
\end{equation} 
and 
\begin{equation}
(\Delta({\hat p}))_{\alpha,\beta} = \frac{\rm i}{\sqrt{2}} (\sigma_\mu d_\mu({\bf p}) \sigma_2 )_{\alpha,\beta}
\end{equation}
is the pair-field, and $s$ is an imaginary time. Hereafter, the notation $d_\mu({\bf p})=A_{\mu,i} {\hat p}_i
$ will be often used. 

First, the GL free energy density in the globally isotropic ($\delta_u=0$) case obtained within the mean field approximation is reviewed. Based on the model (\ref{impline}), the wave number dependences of the scattering amplitude are not considered in $\delta_u=0$ case so that the impurity-induced renormalization of the pair-field vertex (see Fig.1) is not introduced. Then, the GL free energy density $F^{({\rm wc})}=F_2^{({\rm wc})}+F_4^{({\rm wc})}$ can simply be written in the form 
\begin{equation}\label{S2}
F_2^{({\rm wc})}(0) = \biggl[ \frac{\delta_{i,j}}{3 |g|} - T \sum_{\varepsilon} \sum_{{\bf p},{\bf p}'} {\hat p}_i {\hat p}'_j {\overline {{\cal G}_{\varepsilon} ({\bf p}, {\bf p}') \, {\cal G}_{-\varepsilon} (-{\bf p}, -{\bf p}')}} \biggr] A_{\mu,i}^* A_{\mu,j}, 
\end{equation}
with 
\begin{eqnarray}\label{S4wc} 
F_{4}^{({\rm wc})}(0) &\simeq& T \sum_{\varepsilon, {\bf p}} (\, {\cal G}_{\varepsilon}({\bf p}) \, {\cal G}_{-\varepsilon}(-{\bf p}) \,)^2 {\rm Tr}(\Delta_{\hat p}^\dagger \Delta_{\hat p} \Delta^\dagger_{\hat p} \Delta_{\hat p}) 
\nonumber \\
&=& \beta^{({\rm wc})}_1(0) |A_{\mu,i} A_{\mu,i}|^2 + \beta^{({\rm wc})}_2(0) (A_{\mu,i}^* A_{\mu,i})^2 + \beta^{({\rm wc})}_3(0) A^*_{\mu,i} A^*_{\nu,i} A_{\mu,j} A_{\nu,j} \nonumber \\
&+& \beta^{({\rm wc})}_4(0) A^*_{\mu,i} A_{\nu,i} A^*_{\nu,j} A_{\mu,j} + \beta^{({\rm wc})}_5(0) A^*_{\mu,i} A_{\nu,i} A^*_{\mu,j} A_{\nu,j}, 
\end{eqnarray}
where $\beta^{({\rm wc})}_2(0) = \beta^{({\rm wc})}_3(0) = \beta^{({\rm wc})}_4(0) = - \beta^{({\rm wc})}_5(0) = -2 \beta^{({\rm wc})}_1(0)= 2 \beta^{({\rm wc})}(T)$, ${\tilde \varepsilon}=\varepsilon + {\rm sgn}(\varepsilon)/(2 \tau_o)$, 
\begin{equation}\label{betawc}
\beta^{({\rm wc})}(T) = - {\overline \beta}_0(T) \psi^{(2)}(y), 
\end{equation}
${\overline \beta}_0(T) = N(0)/(480 \pi^2 T^2)$, and $y=(4 \pi \tau_o T)^{-1} + 1/2$. Here, ${\cal G} ({\bf p}, {\bf p}')$ is the Green's function defined prior to the random average, and $\psi^{(2)}(z) = - 2 \sum_{n \geq 0}(n + z)^{-3}$ is the second derivative of the digamma function $\psi(z)$. 

Next, the above expressions will be extended to the anisotropic case with $\delta_u \neq 0$. Up to O($\delta_u$), its quadratic term is expressed by 
\begin{equation}
\frac{F_{2}^{({\rm wc})}(\delta_u)}{V N(0)} = \alpha_z^{({\rm wc})} \, A_{\mu,z}^{\ast}A_{\mu,z} + \alpha^{({\rm wc})} \, A_{\mu,j}^{\ast}A_{\mu,j}, 
\label{eq:quadr}
\end{equation}
where 
\begin{eqnarray}
\alpha^{({\rm wc})} &=& \frac{1}{3} \biggl(\ln\frac{T}{T_{c0}}+\psi(y)-\psi\big(\frac{1}{2}\big) + \frac{\delta_u}{4\pi T\tau_o}\frac{1}{5}\psi^{(1)}(y) \biggr), \nonumber \\
\alpha_z^{({\rm wc})} &=& \frac{\delta_u}{4\pi T\tau_o}\frac{16}{45}\psi^{(1)}(y). 
\end{eqnarray}
Here, the $\delta_u$-dependences arise from both the self energy term in ${\cal G}$ and the vertex corrections drawn in Fig.1. 
\begin{figure}[t]
\scalebox{0.6}[0.6]{\includegraphics{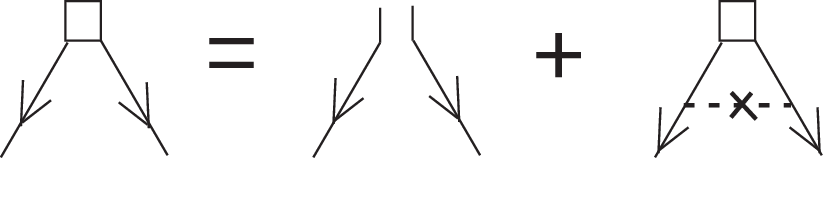}}
\caption{Diagram expressing the impurity-induced vertex corrections. According to eq.(\ref{impline}), the vertex correction expressed by the dashed line with a cross carries the anisotropy parameter $\delta_u$. The solid line with an arrow denotes the quasiparticle Green's function defined in the normal state. }
\label{fig.1}
\end{figure}

The corresponding quartic term $F_4(\delta_u)$ is also derived in a similar manner, and its form is unaffected even if taking account of the SC corrections to be included later. Up to O($\delta_u$), it takes the form 
\begin{widetext}
\begin{eqnarray}
&F_{4}(\delta_u)& \, = \beta_1(\delta_u) |A_{\mu,i}A_{\mu,i}|^2+\beta_2(\delta_u) (A_{\mu,i}^{\ast}A_{\mu,i})^2 
+ \beta_3(\delta_u) A_{\mu,i}^{\ast}A_{\nu,i}^{\ast}A_{\mu,j}A_{\nu,j} 
+ \beta_4(\delta_u) A_{\mu,i}^{\ast}A_{\nu,i}A_{\nu,j}^{\ast}A_{\mu,j} \nonumber \\
&+& \beta_5(\delta_u) A_{\mu,i}^{\ast}A_{\nu,i}A_{\nu,j}A_{\mu,j}^{\ast} 
+ [ \, ( \, \beta_{1z}(\delta_u)  A_{\mu,i}A_{\mu,i}A_{\nu,z}^{\ast}
A_{\nu,z}^{\ast} 
+ \beta_{2z}(\delta_u) A_{\mu,i}^{\ast}A_{\mu,i}A_{\nu,z}^{\ast}A_{\nu,z} 
+ \beta_{3z}(\delta_u) A_{\mu,i}A_{\nu,i}A_{\mu,z}^{\ast}A_{\nu,z}^{\ast} \nonumber \\
&+& \beta_{4z}(\delta_u) A_{\mu,i}^{\ast}A_{\nu,i}A_{\nu,z}^{\ast}A_{\mu,z} 
+ \beta_{5z}(\delta_u) A_{\mu,i}^{\ast}A_{\nu,i}A_{\mu,z}^{\ast}A_{\nu,z} \, ) + {\rm c.c.} \, ]. 
\label{eq:quatrfr}
\end{eqnarray}
\end{widetext}
Each of the coefficients $\beta_i$ ($\beta_{iz}$) is the sum of a weak coupling contribution $\beta_i^{({\rm wc})}$ ($\beta_{iz}^{({\rm wc})}$) and a SC one $\beta_i^{({\rm sc})}$ ($\beta_{iz}^{({\rm sc})}$) to be presented later. 
The coefficients $\beta^{({\rm wc})}_i$ and $\beta_{iz}^{({\rm wc})}$ 
are given by 
\begin{eqnarray}
\beta^{({\rm wc})}_3(\delta_u) &=& -2\beta^{({\rm wc})}_1(\delta_u) = - \, 2 {\overline \beta}_0(T) \Big[\psi^{(2)}(y) 
+ \frac{\delta_u}{4\pi T\tau_o}\frac{1}{7} \psi^{(3)}(y) \Big], \nonumber \\
\beta^{({\rm wc})}_2(\delta_u) &=& \beta^{({\rm wc})}_4(\delta_u) =-\beta^{({\rm wc})}_5(\delta_u) = \beta^{({\rm wc})}_3(\delta_u) - \frac{1}{2\pi T\tau_o} {\overline \beta}_0(T) \Big[\big(\frac{5}{18}+\frac{\delta_u}{54}\big)\psi^{(3)}(y) + \frac{\delta_u}{4\pi T\tau_o} \frac{1}{18} \psi^{(4)}(y) \Big], \nonumber\\
\beta_{3z}^{({\rm wc})}(\delta_u) &=& - 2\beta_{1z}^{({\rm wc})}(\delta_u) = - \, \frac{\delta_u}{2\pi T\tau_o} {\overline \beta}_0(T) \frac{46}{63}\psi^{(3)}(y), \nonumber \\
\beta_{2z}^{({\rm wc})}(\delta_u) &=& \beta_{4z}^{({\rm wc})}(\delta_u) =-\beta_{5z}^{({\rm wc})}(\delta_u) = \beta_{3z}^{({\rm wc})}(\delta_u) - \frac{\delta_u}{2\pi T\tau_o} {\overline \beta}_0(T) \Big[\frac{1}{9}\psi^{(3)}(y) + \frac{1}{4\pi T\tau_o} \frac{4}{27} \psi^{(4)}(y) \Big] 
\label{betawcbulk}
\end{eqnarray}
up to O($\delta_u$). All of the expressions given above have been derived in the previous work \cite{AI06}, although the present notation is slightly different from the previous one. 

In the ensuing sections, the corresponding SC contributions to $\alpha^{({\rm wc})}$ and $\alpha_z^{({\rm wc})}$ need to be derived together with $\beta_i^{({\rm sc})}$ and $\beta_{iz}^{({\rm sc})}$. 

\section{Spin-Fluctuation Model of Strong Coupling Contribution}

As a simple model, we will use the spin-triplet pairing interaction between quasiparticles stemming from the bare interaction of Stoner type 
$H_{\rm bareint}=I \int_{\bf r} {\hat n}_\alpha (\sigma_x)_{\alpha,\beta} {\hat n}_\beta /2$, where ${\hat n}_\alpha$ is the bare fermion density with the spin projection $\alpha$. This bare interaction results in the ferromagnetic spin critical fluctuation (paramagnon) \cite{Nakajima}. If this paramagnon is treated in the Gaussian approximation so that any mode-couplings between the paramagnons are neglected, the resulting free energy $F^{(s)}$ corresponding to that of the free paramagnon is given by 
\begin{equation}
F^{(s)} = \frac{T}{2} \sum_{\mu,\nu} \sum_{\omega} \int_{\bf q} [ {\rm ln}(1 - I \chi_{\mu,\nu}({\bf q},\omega)) + I \chi_{\mu,\nu}({\bf q}, 
\omega)],
\end{equation}
where $\int_{\bf q} = \int d^3{\bf q}/(2 \pi)^3$, and the dynamical susceptibility is expressed in the form 
\begin{equation}
\chi_{\mu,\nu}({\bf q},\omega) = - \frac{T}{2} \sum_{\varepsilon, {\bf p}, {\bf p}'} (\sigma_\mu)_{\alpha, \beta} (\sigma_\nu)_{\gamma, \delta} [ {\overline {{\cal G}_{\beta, \gamma}({\bf p}_-, {\bf p}'_-; \varepsilon) {\cal G}_{\delta, \alpha}({\bf p}'_+, {\bf p}_+; \varepsilon+\omega)}} - {\overline {{\cal F}_{\beta, \delta}({\bf p}_-, {\bf p}'_-; \varepsilon) {\cal F}^\dagger_{\gamma,\alpha}({\bf p}'_+, {\bf p}_+; \varepsilon+\omega)}} ]
\end{equation}
when the interaction Hamiltonian of quasiparticles takes the quadratic form (\ref{bcsint}) according to the mean field approximation, even in the presence of an impurity disorder, and ${\cal F}$ and ${\cal F}^\dagger$ are the anomalous Matsubara Green's functions. When $F^{(s)}$ is expanded in powers of the difference 
\begin{equation}
\delta \chi_{\mu,\nu}({\bf q}, \omega) = \chi_{\mu, \nu}({\bf q}, \omega) - \chi_{\mu,\nu}({\bf q}, \omega)|_{\Delta=0}, 
\end{equation}
its lowest order term is \cite{correc} 
\begin{equation}
F^{(s)}_2 = \frac{- 4 T k_{\rm F}^3}{N(0)} \sum_\mu \sum_\omega \int\frac{dq q^2}{(2 \pi)^2} \, \langle \delta \chi_{\mu,\mu}({\bf q}, \omega) \rangle_{\hat q}  \frac{{\overline I}^2}{1 - {\overline I} + {\overline q}^2{\overline I}/3 + \pi |\omega|/(8 E_{\rm F} {\overline q})}, 
\label{paramag2}
\end{equation}
where $\langle \delta \chi_{\mu,\mu}({\bf q}, \omega) \rangle_{\hat q}$ is the average of $\delta \chi_{\mu,\nu}({\bf q}, \omega)$ on the orientation of ${\bf q}$, ${\overline q}=q/(2k_{\rm F})$, and $q=|{\bf q}|$. 
According to the Brinkman et al.(BSA) \cite{BSA}, $F^{(s)}$ will be treated as follows: Up to O($|\Delta|^2$) and with no impurity-induced vertex correction, we have 
\begin{equation}
\delta \chi_{\mu,\nu}({\bf q}, \omega) 
= \frac{T}{2} \sum_{\varepsilon, {\bf p}} [{\rm Tr}(\sigma_\mu \Delta^{\rm T}_{\bf p} \sigma^{\rm T}_\nu \Delta^*_{\bf p}) |{\cal G}_{{\bf p}_-}(\varepsilon)|^2 |{\cal G}_{{\bf p}_+}(\varepsilon+\omega)|^2 + 2 {\rm Tr}(\sigma_\mu \sigma_\nu \Delta_{\bf p}\cdot\Delta^\dagger_{\bf p}) {\cal G}_{{\bf p}_+}(\varepsilon - \omega) {\cal G}_{{\bf p}_-}(-\varepsilon) ({\cal G}_{-{\bf p}_-}(\varepsilon))^2].
\label{deltachi}
\end{equation}
After integrating this expression over $\xi$ which is the kinetic energy measured from the Fermi energy, the denominator $1/[({\bf v}_{\bf p}\cdot{\bf q})^2 + \omega^2]$ appears, where ${\bf v}_{\bf p}$ is the velocity $\parallel {\bf p}$. The denominator of the paramagnon propagator seen in eq.(\ref{paramag2}) implies that the $q$-integral is dominated in the region ${\overline q} \simeq \sqrt{1 - {\overline I}}$, and $|\omega| < E_{\rm F} \sqrt{1 - {\overline I}}$, suggesting that ${\rm Max}(|\varepsilon|, |\omega|)/(v_{\rm F}q)$ can be regarded as a small parameter. For this reason, we consider the average of the denominator $\langle 1/[({\bf v}_{\bf p}\cdot{\bf q})^2 + \omega^2] \rangle_{\hat {\bf q}}$ and replace this denominator with
\begin{equation}
\delta_{{\bf p}\cdot{\bf q},0} \biggl( \frac{\pi}{2 v_{\rm F}|q||\omega|} - \frac{1}{v_{\rm F}^2 q^2} \biggr) 
\label{deltaden}
\end{equation}
in order for the $|\omega|/(v_{\rm F}q)$-expansion of $\langle 1/[({\bf v}_{\bf p}\cdot{\bf q})^2 + \omega^2] \rangle_{\hat {\bf q}}$ to be recovered 
properly. The second term of eq.(\ref{deltaden}) which has not been taken into account in Ref.\cite{BSA} will also be included hereafter. 
Then, eq.(\ref{deltachi}) is expressed in the form 
\begin{widetext}
\begin{eqnarray}
\delta \chi_{\mu,\nu}({\bf q}, \omega)
&=& \frac{\pi^2 T N(0)}{4 v_{\rm F} |q|}  \langle {\hat p}_i {\hat p}_j \delta_{{\bf p}\cdot{\bf q},0} \rangle_{\hat {\bf p}} \biggl[ - A_{\rho,i}^* A_{\rho,j} \delta_{\mu,\nu} \sum_\varepsilon {\rm sgn}(\varepsilon) {\rm sgn}(\varepsilon-\omega) \biggl(\frac{1}{{\tilde \varepsilon}^2}+\frac{1}{|{\tilde {\varepsilon - \omega}}|^2} \biggr) \nonumber \\ 
&+& \sum_\varepsilon \frac{2}{|{\tilde \varepsilon}||{\tilde {\varepsilon - \omega}}|} (\delta_{\mu,\nu} A_{\rho,i}^* A_{\rho,j} - A_{\mu,i}^* A_{\nu,j} - A_{\nu,i}^* A_{\mu,j}) 
+ \frac{4}{\pi v_{\rm F} |{\bf q}|} \sum_\varepsilon \frac{1}{|{\tilde \varepsilon}|} (A_{\mu,i}^* A_{\nu,j} + A_{\nu,i}^* A_{\mu,j}) \biggr], 
\label{deltachi2}
\end{eqnarray}
\end{widetext}
where the angle average $\langle {\hat p}_i {\hat p}_j \delta_{{\bf p}\perp{\bf q},0} \rangle_{\hat {\bf p}}$ is given by $\delta^{\rm T}_{i,j}/2$ with $\delta^{\rm T}_{i,j} \equiv \delta_{i,j}-{\hat q}_i {\hat q}_j$. 

If using eq.(\ref{deltachi2}) in $F^{(s)}_2$, some contribution to the $A_{\mu,j}^* A_{\mu,j}$  term is obtained. However, this SC contribution to the quadratic term may be neglected, because this does not lead to distinguishing various pairing states from one another and may be absorbed into a definition of $T_c$. On the other hand, when applying eq.(\ref{deltachi}) to the next order term of $F^{(s)}$ 
\begin{equation}
F_4^{(s)} = - \frac{T (2 k_{\rm F})^3}{8 (N(0)\pi)^2} \sum_{\mu,\nu} \sum_\omega \int_0^1 dq q^2 \biggl(\frac{\overline I}{1 - {\overline I} + {\overline q}^2{\overline I}/3 + \pi |\omega|/(8 E_{\rm F} {\overline q})} \biggr)^2 \langle (\delta \chi_{\mu,\nu}({\bf q}, \omega))^2 \rangle_{\hat q},
\label{quarfreesf}
\end{equation}
the following corrections to the coefficients $\beta_1$ and $\beta_2$ of the GL quartic terms are 
found \cite{AI05}: 
\begin{eqnarray}
\beta_{1,{\rm se}}^{({\rm sc})} &=& \frac{{\overline \beta}_0(T)}{300} \psi^{(2)}\biggl(\frac{1}{2}\biggr) \, t \, \delta \, \sum_m (D_1^{(d)}(m))^2, \nonumber \\ 
\beta_{2,{\rm se}}^{({\rm sc})} &=& \beta_{1,{\rm se}}^{({\rm sc})} \, \biggl[\sum_m (9 (D_2^{(d)}(m))^2 - 6 D_1^{(d)}(m) D_2^{(d)}(m) - 2(D^{(d)}_1(m))^2) \biggr] \, \biggl( \sum_m 
(D_1^{(d)}(m))^2 \biggr)^{-1}, \nonumber \\
\beta_{3,{\rm se}}^{({\rm sc})} &=& \frac{1}{6} (\beta_{2,{\rm se}}^{({\rm sc})} + 5 \beta_{1,{\rm se}}^{({\rm sc})}), \nonumber \\
\beta_{4,{\rm se}}^{({\rm sc})} &=& \beta_{3,{\rm se}}^{({\rm sc})} + 5 \beta_{1,{\rm se}}^{({\rm sc})}, \nonumber \\ 
\beta_{5,{\rm se}}^{({\rm sc})} &=& 7 \beta_{1,{\rm se}}^{({\rm sc})}, 
\label{bulkbetasc}
\end{eqnarray}
where $
\delta$ ($\propto T/E_{\rm F}$) is a parameter defined in Ref.\cite{BSA}, and 
\begin{equation}
D_1^{(d)}(m) = \frac{1}{2} \biggl(\frac{1}{|m|} + \frac{1}{|m| + (2 \pi \tau_o T)^{-1}} \biggr) \, (\psi(y + |m|) - \psi(y)), 
\end{equation}
\begin{equation}
D_2^{(d)}(m)=\frac{1}{2} \psi^{(1)}(y + |m|). 
\end{equation}
These are nothing but the extension in the relaxation time approximation of the BSA results in clean limit to the disordered case. Contrary to the experimental suggestion \cite{HalperinJPSJ,Yoonseok}, however, this relaxation time approximation rather enhances the temperature region in which the A-phase is more stable than the B-phase and thus, is insufficient as a description of the superfluid $^3$He in globally isotropic aerogel \cite{Aoyama}. 

\begin{figure}[t]
\scalebox{0.45}[0.45]{\includegraphics{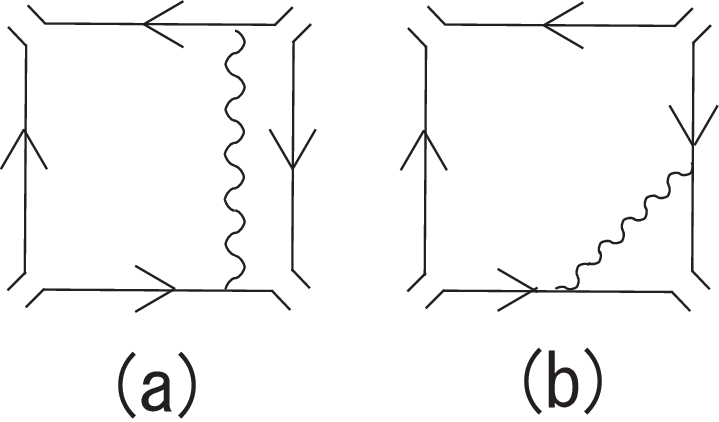}}
\caption{Diagrams ignored previously \cite{BSA,RS} in incorporating the SC corrections to the GL quartic terms for the bulk liquid. The wave line denotes the paramagnon (spin-fluctuation) propagator defined in the normal state.}
\label{fig.2}
\end{figure}

Inclusion of the O($|\Delta|^4$) contribution to $\delta \chi_{\mu,\nu}({\bf q}, \omega)$ in $F_s^{(2)}$ also induces corrections to $\beta_j$. In the previous studies \cite{BSA,RS}, the quartic terms following from $F_s^{(2)}$ have been neglected. This omission has been justified there based on the use of the static approximation for the four-point vertex $\Gamma_4$, i.e., on the neglect of the frequency dependence of $\Gamma_4$ which, in the present paramagnon approach, corresponds to the normal paramagnon propagator. For instance, the diagram of Fig.2 (a) under this static approximation becomes identically zero. Further, Fig.2 (b) has been interpreted as being absorbed into the vertex correction of the type indicated in Fig.1. Hereafter, regarding the results in clean limit, we will follow the interpretation in previous works \cite{BSA,RS}. 

When trying to incorporate vertex corrections induced by the impurity-scattering effects due to the aerogel structures, however, the SC corrections reflected in $F_s^{(2)}$ are found to play important roles. 
In a previous study, impurity-induced vertex corrections to the SC effect on the quartic terms have been examined \cite{Aoyama} based on the static approximation in the work \cite{RS}, in order to verify the suggestion from experiments that the SC correction is significantly suppressed in aerogel possibly with no global anisotropy \cite{Yoonseok}. It has been found that the impurity-induced vertex correction qualitatively reduces the SC correction, although it is not substantial quantitatively. However, important impurity-induced terms of the SC contribution to $\beta_j$, created from the self energy correction due to the paramagnon in clean limit, have been overlooked in Ref.\cite{Aoyama}. These terms, sketched in Fig.3, become more important at higher values of the frequency carried by the paramagnon propagator, and this is why these terms have not been examined in the approach based on the static approximation \cite{RS}. At least within the conventional paramagnon approach \cite{BSA}, these terms are dominant contributions to the impurity-induced SC correction of O($1/(\tau_o T)$) and result in a significant reduction of the SC correction and, hence, in the absence of the A-phase in globally isotropic aerogels \cite{HalperinJPSJ,Yoonseok}. 

Detailed expressions on the terms depicted in Fig.3 will be given here. For instance, the contribution to $\delta \chi_{\mu,\nu}({\bf q},\omega)$ of Fig.3 (a) is 
\begin{equation}
2 \times {\rm Fig.3 (a)} = - \frac{T}{4 \pi N(0) \tau_o}  \delta_{\mu,\nu} \sum_\varepsilon \biggl[ \int_{\bf p} |{\bf d}({\bf p})|^2  {\cal G}_{{\bf p}+{\bf q}}(\varepsilon+\omega) [{\cal G}_{\bf p}(\varepsilon)]^2 {\cal G}_{-{\bf p}}(-\varepsilon) \biggr]^2.  
\label{isochi(1)}
\end{equation}
After performing the $\xi_{\bf p}$-integral, the expression in the square bracket of eq.(\ref{isochi(1)}) becomes 
\begin{equation}
\frac{\pi N(0)}{2} \frac{|\varepsilon|+|\varepsilon+\omega|}{\varepsilon^2} \biggl\langle \frac{|{\bf d}({\bf p})|^2}{({\bf v}\cdot{\bf q})^2 + (|\varepsilon|+|\varepsilon+\omega|)^2}\biggr\rangle_{\hat p}. 
\label{isochi(1)var}
\end{equation}
Using the BSA's procedure for the ${\hat p}$-average, the contribution of eq.(\ref{isochi(1)}) to $\langle \delta \chi_{\mu,\nu}({\bf q},\omega) \rangle_{\hat q}$ becomes 
\begin{equation}
- \frac{3 \pi^4}{80} \frac{T N(0)}{4 \pi \tau_o v_{\rm F}^2 q^2} \delta_{\mu,\nu} \sum_{\varepsilon > 0} \frac{1}{\varepsilon^4} \biggl(\delta_{i,j} \delta_{k,l} + \frac{1}{6} (\delta_{i,k} \delta_{j,l} + \delta_{i,l} \delta_{j,k}) \biggr)  A_{\rho, i}^* A_{\rho, j} A_{\lambda, k}^* A_{\lambda, l}. 
\end{equation}

\begin{figure}[t]
\scalebox{0.5}[0.5]{\includegraphics{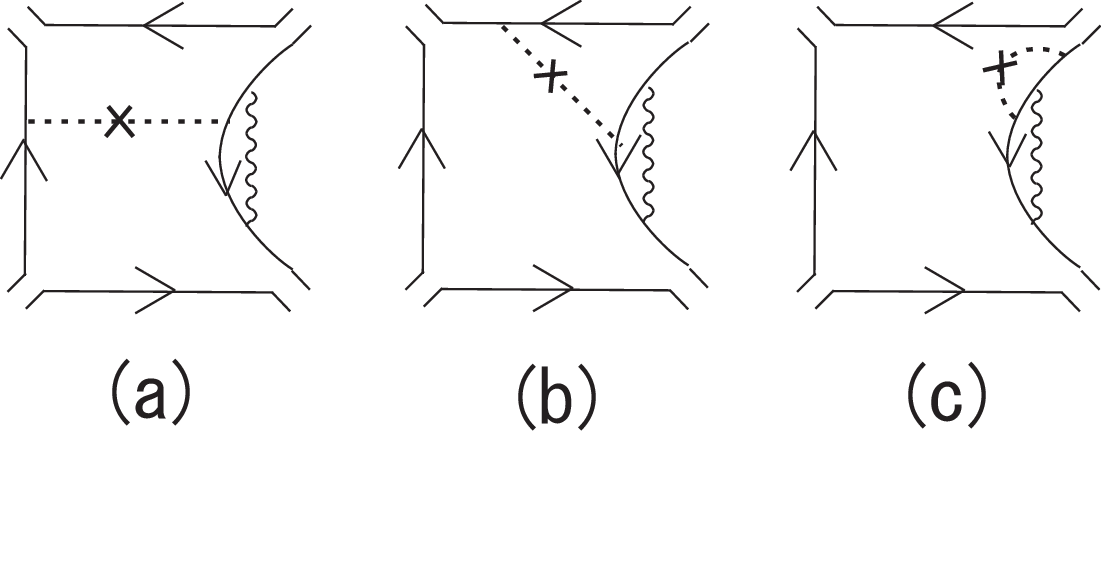}}
\caption{SC contributions to the GL quartic terms playing important roles in a globally isotropic disordered system. Here, the dashed curve denotes eq.(\ref{impline}). }
\label{fig.3}
\end{figure}

As a result of the use of the BSA's procedure, the $\omega$-dependence has been lost, implying that the corresponding contribution to the free energy inevitably depends on the high energy cutoff $E_c$ of the spin-fluctuation dynamics. It should be note that this dependence does not seem to be an artifact of the BSA procedure: The expression (\ref{isochi(1)var}) suggests that the $|\omega|^{-1}$-dependence at large $|\omega|$ appears only in the low $|{\bf q}|$-contribution which, due to the $|\omega|/q$-dependence of the normal paramagnon propagator, is not dominant in the ${\bf q}$-integral for obtaining the free energy. In this manner, we judge that the SC correction from Fig.3 (a) is inevitably dominated by the high $|\omega|$-contributions and is practically dependent on the high energy cutoff $E_c$ ($< E_{\rm F}\sqrt{1 - {\overline I}}$). In the same way, the contributions of Fig.3 (b) and (c) to $\langle \delta \chi_{\mu,\nu}({\bf q},\omega) \rangle_{\hat q}$ become 
\begin{eqnarray}
4 &\times& {\rm Fig.3 (b)} = - \frac{T N(0) \pi^2}{10 \pi \tau_o v_{\rm F}^2 q^2} \delta_{\mu,\nu} \sum_{\varepsilon > 0} \frac{1}{\varepsilon^4} (\delta_{i,j} \delta_{k,l} + \delta_{i,k} \delta_{j,l} + \delta_{i,l} \delta_{k,j}) A_{\rho, i}^* A_{\rho, j} A_{\lambda, k}^* A_{\lambda, l}, \nonumber \\
4 &\times& {\rm Fig.3 (c)} = \frac{3 T N(0) \pi^2 \delta_{\mu,\nu}}{160 \pi \tau_o v_{\rm F}^2 q^2}  (1 - \delta_{\omega, 0}) \sum_{\varepsilon > 0} \biggl(\frac{1}{(\varepsilon+|\omega|)^4} 
- \frac{1}{\varepsilon^4} \biggr) (\delta_{i,j} \delta_{k,l} + \delta_{i,k} \delta_{j,l} + \delta_{i,l} \delta_{k,j}) A_{\rho, i}^* A_{\rho, j} A_{\lambda, k}^* A_{\lambda, l},
\label{isochi(23)}
\end{eqnarray}
respectively. It is clear that both of eq.(\ref{isochi(23)}) are also dominated by the high frequency contributions. 

Finally, the corresponding contributions of Fig.3 to the $\beta_j$-parameters, $\beta_{j, {\rm vc}}^{({\rm sc})}(0)$, will be given here : 
\begin{eqnarray}
\beta_{2, {\rm vc}}^{({\rm sc})}(0) &=& 30 \frac{(\pi {\overline I})^2}{2 \pi \tau_o T} {\overline \beta}_0 \sum_m D_1(|m|) \biggl[ \biggl(\frac{2}{15} + \frac{\pi^2}{80} \biggr) \psi^{(3)}(y) + \frac{\pi^2}{40} (\psi^{(3)}(y) - \psi^{(3)}(y+|m|) \biggr], \nonumber \\
\beta_{3, {\rm vc}}^{({\rm sc})}(0) &=& \beta_{4, {\rm vc}}^{({\rm sc})}(0) 
= 30 \frac{(\pi {\overline I})^2}{2 \pi \tau_o T} {\overline \beta}_0 \sum_m D_1(|m|) \biggl[ \biggl(\frac{2}{15} + \frac{\pi^2}{480} \biggr) \psi^{(3)}(y) + \frac{\pi^2}{240}(\psi^{(3)}(y) - \psi^{(3)}(y+|m|)) \biggr],
\label{isobetasc}
\end{eqnarray}
and $\beta_{1, {\rm vc}}^{({\rm sc})}(0) = \beta_{5, {\rm vc}}^{({\rm sc})}(0)=0$, where 
\begin{equation}
D_1(|m|) = \frac{T}{8 \pi^2 E_{\rm F}} \int_0^\infty d{\overline q} \biggl( 1 - {\overline I} + \frac{\overline I}{3} {\overline q}^2 + \frac{\pi^2 T}{4 |{\overline q}| E_{\rm F}} |m| \biggr)^{-1}. 
\end{equation} 
These corrections to $\beta_j$ are positive so that stability of the A-phase is diminished. Then, for reasonable $\tau_o^{-1}$ values used in our numerical calculations and the ${\overline I}$-values appropriate for the bulk liquid, we find that, at least within the present mean field analysis, the B phase is the only superfluid phase in the globally isotropic case and that the A-phase is not realized anywhere at least below 30 bar. 

\section{Anisotropic Strong Coupling Contributions}

As mentioned in the preceding section, the SC correction to the quadratic term of the GL free energy is negligible even in the presence of impurity-scattering effects as far as the medium is globally isotropic. In contrast, in the case with a global anisotropy, the SC correction to the quadratic term is no longer negligible and, as seen below, plays important roles in determining the pressure dependence of the superfluid transition. The corresponding SC effects on the quartic GL terms will be discussed in the last half of this section. 

Hereafter, we focus on the lowest order contributions in the anisotropy parameter $\delta_u$ to the SC corrections to the GL free energy terms. Then, the parameter $\delta_u$ is carried by a single impurity line appearing as a vertex correction or by a quasiparticle damping of a Green's function in a diagram expressing $\delta \chi_{\mu,\nu}({\bf q}, \omega)$. 

\begin{figure}[b]
\scalebox{0.4}[0.4]{\includegraphics{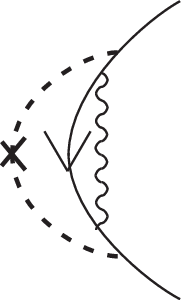}}
\caption{Example of self energy terms regarded in the present study as being absorbed into the weak coupling terms.}
\label{fig.4}
\end{figure}

 First, let us discuss the anisotropic contribution to $\delta \chi_{\mu,\nu}({\bf q}, \omega)$ due to the quasiparticle damping. 
This contribution to $\delta \chi_{\mu,\nu}$ is easily obtained from the corresponding expression in clean limit simply by adding $\eta_{\bf p}$ to $|\varepsilon|$, $|\varepsilon - \omega|$. However, we will not incorporate such a self energy diagram that can be regarded as being absorbed into a weak-coupling process. Its example is raised in Fig.4. The contribution to $\delta \chi_{\mu,\nu}$ accompanied by the anisotropy parameter $\delta_u$ in the self energy correction becomes 
\begin{widetext}
\begin{eqnarray}
\delta \chi^{(1)}_{\mu,\nu}({\bf q},\omega)|_{\rm se} &=& \pi^2 \frac{N(0)}{v_{\rm F}|{\bf q}|} \frac{\delta_u}{\tau_o} \langle {\hat p}_i {\hat p}_j {\hat p}_z^2 \delta_{{\bf p}\perp{\bf q},0} \rangle_{\hat {\bf p}} \biggl[ \frac{T}{2} \sum_{\varepsilon > 0} \frac{1}{(\varepsilon + |\omega|)^3} \delta_{\mu,\nu} A_{\rho,i}^* A_{\rho,j} - T \sum_{\varepsilon > 0}\biggl( \delta_{\omega,0} \frac{1}{\varepsilon^3} + (1 - \delta_{\omega,0}) \biggl[ \frac{1}{|\omega| \varepsilon(\varepsilon+|\omega|)} \nonumber \\
&+& \frac{1}{2}\biggl(1 + \frac{\tau_o|\omega|}{1 + \tau_o|\omega|} \biggr) \biggl(\frac{1}{\varepsilon}+\frac{1}{\varepsilon+|\omega|} \biggr) \frac{1}{\varepsilon(\varepsilon+|\omega|)} \biggr] \biggr) (\delta_{\mu,\nu} A_{\rho,i}^* A_{\rho,j} - A_{\mu,i}^* A_{\nu,j} - A_{\nu,i}^* A_{\mu,j}) \biggr]
\label{deltachiani2}
\end{eqnarray}
\end{widetext}
up to O($\delta_u$), where 
${\tilde \varepsilon}={\rm sgn}(\varepsilon)(|\varepsilon|+1/(2 \tau_o))$, and the angle average $\langle {\hat p}_i {\hat p}_j {\hat p}_z^2 \delta_{{\bf p}\perp{\bf q},0} \rangle_{\hat {\bf p}}$ is given by $[(1 - {\hat q}_z^2) \delta^{\rm T}_{i,j} + 2 \delta^{\rm T}_{i,z} \delta^{\rm T}_{z,j}]/8$. Performing the ${\hat {\bf q}}$-average and substituting it into eq.(\ref{paramag2}), the corresponding contributions to the quadratic terms of the GL free energy density are 
\begin{equation}
\frac{F_s^{(2)}|_{\rm se}}{N(0) V} = \alpha_z^{({\rm sc})}|_{\rm se} \, \, A_{\mu,z}^* A_{\mu,z} + \alpha^{({\rm sc})}|_{\rm se} \, \, A_{\mu,j}^* A_{\mu,j}
\label{aniquadrsc}
\end{equation} 
with $\alpha^{({\rm sc})}|_{\rm se} = \alpha_z^{({\rm sc})}|_{\rm se}/2 = \Sigma_2^{({\rm ss})} + \Sigma_2^{({\rm sv})}$, where 
\begin{widetext}
\begin{eqnarray}
\Sigma_2^{({\rm ss})} &=& \frac{\pi^2 {\overline I}^2}{15} \frac{\delta_u}{4 \pi T \tau_o} \biggl[ \frac{1}{4} D_2(0) \psi^{(2)}(y) + \sum_{m \geq 1} D_2(|m|) \biggl(\frac{3}{2} \psi^{(2)}(y + m) + \frac{2}{(m+1/(2 \pi \tau_o T))^2} ( \psi(y + m) - \psi(y)) \nonumber \\ 
&-& \biggl(\frac{1}{m} 
+ \frac{1}{m+1/(2 \pi \tau_o T)} \biggr) (\psi^{(1)}(y + m) - \psi^{(1)}(y)) \biggr) \biggr], \nonumber \\
\Sigma_2^{({\rm sv})} &=& \frac{8 \pi^2 {\overline I}^2}{15} \frac{\delta_u}{4 \pi T \tau_o} \frac{1}{4} \sum_{m > 0} D_1(|m|) (7 \psi^{(1)}(y) - 3 \psi^{(1)}(y + m) ),
\label{selfself2}
\end{eqnarray}
\end{widetext}
and
\begin{equation}
D_2(|m|) = \frac{1}{4 \pi^2} \int_0^{1} d{\overline q}^2 \biggl( 1 - {\overline I} + \frac{\overline I}{3} {\overline q}^2 + \frac{\pi^2 T}{4 |{\overline q}| E_{\rm F}} |m| \biggr)^{-1} 
\end{equation}
Here, the $\Sigma_2^{({\rm sv})}$-term results from the contributions to $\delta \chi^{(1)}_{\mu,\nu}$ of the second term of eq.(\ref{deltaden}), which are not shown in eq.(\ref{deltachiani2}). 
In general, these self energy contributions to the quadratic term have the same sign as the corresponding weak-coupling term and, for instance, widen the region of the polar phase at lower pressures in $^3$He in a stretched aerogel. However, these contributions are overcome at higher pressures by the vertex correction contributions given below which have the opposite sign to that of the weak-coupling term. 

\begin{figure}[t]
\scalebox{0.45}[0.45]{\includegraphics{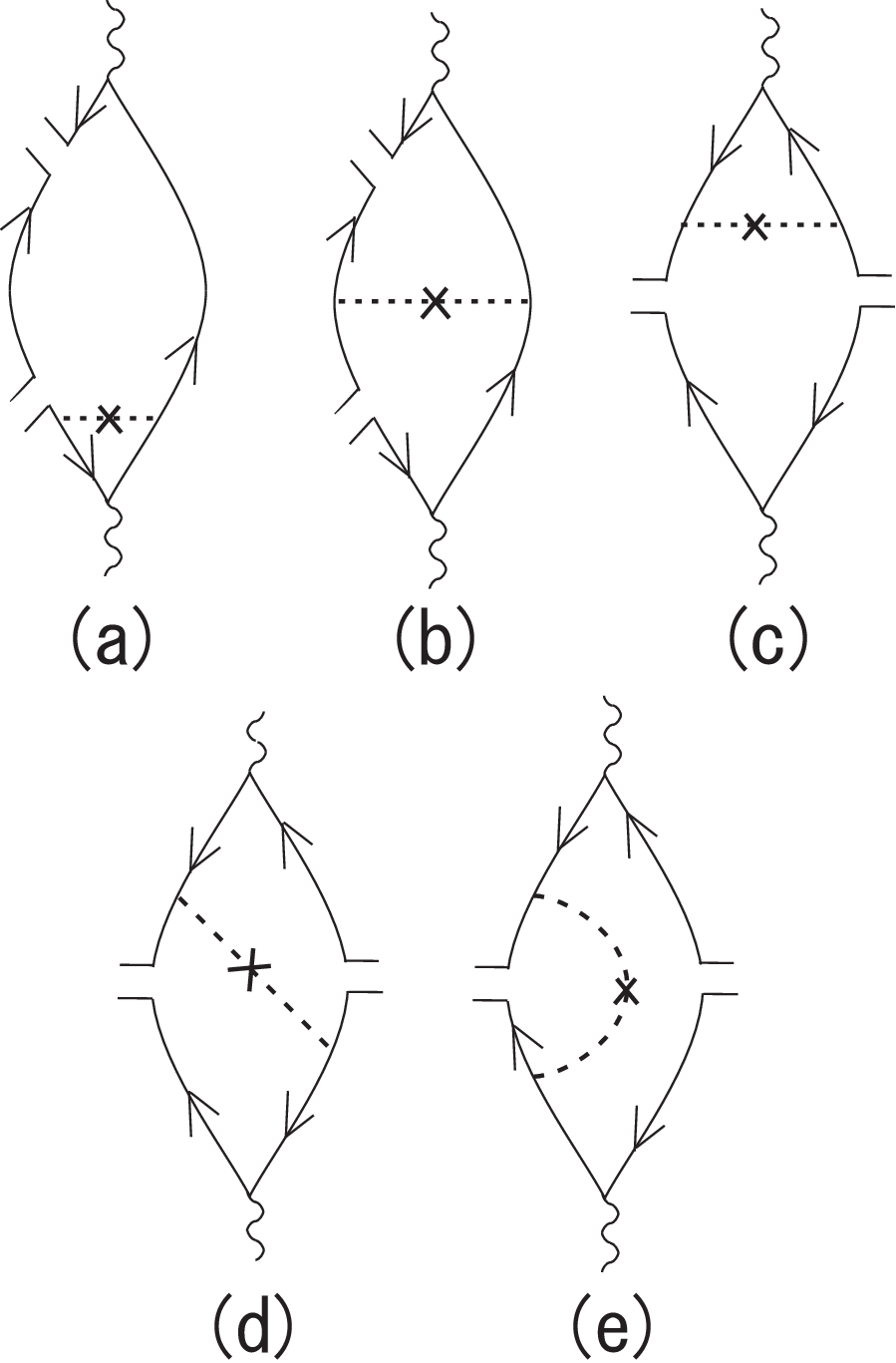}}
\caption{Diagrams contributing to $\delta \chi_{\mu, \nu}$ with an impurity-induced vertex correction. Here, the impurity (dashed) line carries the parameter $\delta_u$. }
\label{fig.5}
\end{figure}

Next, let us turn to the anisotropic contributions to $\delta \chi_{\mu,\nu}({\bf q}, \omega)$ accompanied by impurity-induced vertex corrections. Related diagrams are shown in Fig.5. Like in the self energy contributions, the diagrams to be regarded as being absorbed into weak-coupling ones will be neglected. 
The contribution of each diagram to $\delta \chi_{\mu,\nu}({\bf q},\omega)/N(0)$ is given by 
\begin{widetext}
\begin{eqnarray}
4 \times {\rm Fig.5 (a)} &=& \frac{\pi^4}{4} \frac{T}{v_{\rm F}^2 q^2} \sum_{\varepsilon} \theta(\varepsilon(\omega - \varepsilon)) \biggl(\frac{1}{\varepsilon^2} + \frac{1}{(\varepsilon - \omega)^2} \biggr) \biggl\langle \biggl\langle \delta_{{\bf p}\perp{\bf q},0} \delta_{{\bf p}'\perp{\bf q},0} \frac{\delta_u}{2 \pi \tau_o N(0)}({\hat {\bf p}}_z-{\hat {\bf p}'}_z)^2 {\rm Tr}(\sigma_\mu \sigma_\nu \Delta_{\bf p}\cdot\Delta^\dagger_{\bf p}) \biggr\rangle_{\hat {\bf p}} \biggr\rangle_{\hat {\bf p}'}, \nonumber \\
2 \times {\rm Fig.5 (b)} &=& - \frac{\pi^4}{4} \frac{T}{v_{\rm F}^2 q^2} \sum_\varepsilon \frac{1}{\varepsilon^2} \biggl\langle \biggl\langle \delta_{{\bf p}\perp{\bf q},0} \delta_{{\bf p}'\perp{\bf q},0} \frac{\delta_u}{2 \pi \tau_o N(0)}({\hat {\bf p}}_z-{\hat {\bf p}'}_z)^2 {\rm Tr}(\sigma_\mu \sigma_\nu \Delta_{{\bf p}'}\cdot\Delta^\dagger_{\bf p}) \biggr\rangle_{\hat {\bf p}} \biggr\rangle_{\hat {\bf p}'}, \nonumber \\
2 \times {\rm Fig.5 (c)} &=& \frac{\pi^4}{2} \frac{T}{v_{\rm F}^2 q^2} \sum_\varepsilon \theta(\varepsilon(\omega - \varepsilon)) \frac{1}{|\varepsilon(\omega - \varepsilon)|} \biggl\langle \biggl\langle \delta_{{\bf p}\perp{\bf q},0} \delta_{{\bf p}'\perp{\bf q},0} \frac{\delta_u}{2 \pi \tau_o N(0)}({\hat {\bf p}}_z-{\hat {\bf p}'}_z)^2 {\rm Tr}(\sigma_\nu \Delta_{{\bf p}} \sigma_\mu^{\rm T}\Delta^\dagger_{\bf p}) \biggr\rangle_{\hat {\bf p}} \biggr\rangle_{\hat {\bf p}'}, \nonumber \\
2 \times {\rm Fig.5 (d)} &=& \frac{\pi^4}{8} \frac{T}{v_{\rm F}^2 q^2} \delta_{\omega,0} \sum_\varepsilon \frac{1}{\varepsilon^2} \biggl\langle \biggl\langle \delta_{{\bf p}\perp{\bf q},0} \delta_{{\bf p}'\perp{\bf q},0} \frac{\delta_u}{2 \pi \tau_o N(0)}({\hat {\bf p}}_z-{\hat {\bf p}'}_z)^2 [ {\rm Tr}(\sigma_\nu \Delta_{{\bf p}} \sigma_\mu^{\rm T} \Delta^\dagger_{{\bf p}'} + \sigma_\nu \Delta_{{\bf p}'} \sigma_\mu^{\rm T} \Delta^\dagger_{\bf p}) \biggr\rangle_{\hat {\bf p}} \biggr\rangle_{\hat {\bf p}'}, \nonumber \\
2 \times {\rm Fig.5 (e)} &=& \frac{\pi^3}{4} \frac{T}{v_{\rm F} |q|} \sum_\varepsilon \frac{1}{|\varepsilon(\omega - \varepsilon)|} \biggl(\frac{1}{|\varepsilon|} + \frac{1}{|\omega -\varepsilon|} \biggr) \biggl\langle \biggl\langle \delta_{{\bf p}\perp{\bf q},0} \frac{\delta_u}{2 \pi \tau_o N(0)}({\hat {\bf p}}_z-{\hat {\bf p}'}_z)^2 {\rm Tr}(\sigma_\nu \Delta_{{\bf p}'} \sigma_\mu^{\rm T} \Delta^\dagger_{\bf p}) \biggr\rangle_{\hat {\bf p}} \biggr\rangle_{\hat {\bf p}'}
\end{eqnarray}
\end{widetext}

The last figure plays a similar role to the self energy contributions of eqs.(\ref{selfself2}). On the other hand, the remaining four kinds of diagrams have a crossing between the normal paramagnon and the impurity lines, like those of Fig.3, and hence, are of higher order in $T/E_{\rm F}$. Due to this additional $E_{\rm F}^{-1}$ dependence, they are enhanced with increasing the pressure $P$. 
In particular, the diagrams of Fig.5 (a) and (b) sensitive to the high energy cutoff $E_c$  become dominant at high pressures. 

The crucial feature of these crossing diagrams is that they have the sign competitive with that of the weak-coupling term. This fact leads to a pressure-induced sign reversal on the anisotropy and an appearance of a critical point on $T_c(P)$-curve (see sec.IV). At higher pressures, these SC contributions become more dominant than the competitive weak-coupling ones, and consequently, the superfluid feels the opposite anisotropy to the genuine one determined from the aerogel structure. Thus, for instance, a situation occurs in which the polar phase and A$_{\rm XY}$ one with ${\hat {\bf l}}$-vector perpendicular to the anisotropy axis become unstable in a unaxially stretched aerogel \cite{HalperinNP}. 
Note that the diagrams (a) and (b) in Fig.5 are similar to those in Fig.3 which were the main terms in the SC correction including the {\it isotropic} impurity scattering explained in sec.II. That is, the SC contributions which are negligible in the isotropic case become rather important in the anisotropic cases. 

The contributions of the diagrams in Fig.5 to the free energy density are 
expressed as 
\begin{equation}
\frac{F_s^{(2)}|_{\rm vc}}{N(0) V} = \alpha^{({\rm sc})}_z|_{\rm vc} \, \, A_{\rho,z}^* A_{\rho,z} + \alpha^{({\rm sc})}|_{\rm vc} \, \, 
A_{\rho,j}^* A_{\rho,j} 
\label{aniquadrvc}
\end{equation}
with $\alpha^{({\rm sc})}|_{\rm vc} = - \pi^3 \delta_u {\overline I}^2 \Lambda_\perp/(3 T \tau_o)$, and $\alpha_z^{({\rm sc})}|_{\rm vc} = - \pi^3 \delta_u {\overline I}^2 [ \Lambda_1 - \Lambda_\perp 
+ \Lambda_2 ]/(3 T \tau_o)$, where 
\begin{eqnarray}
\Lambda_1 &=& \frac{3}{2}\biggl[ \biggl(\frac{1}{10}+\frac{8}{9 \pi^2} \biggr) D_1(0) \psi^{(1)}(y) + \sum_{m \geq 1} D_1(m) \biggl[ \biggl(\frac{4}{3}+\frac{8}{9 \pi^2} \biggr) \psi^{(1)}(y) - \psi^{(1)}(y + m) \nonumber \\
&+& \frac{1}{m} \biggl(\frac{1}{3}+\frac{16}{9 \pi^2} \biggr) (\psi(y+ m) - \psi(y) ) \biggr] \biggr], \nonumber \\
\Lambda_2 &=& - \frac{1}{2 \pi^2} \biggl[ \frac{1}{6} D_2(0) \psi^{(2)}(y) + \sum_{m \geq 1} D_2(m) \bigg[ \frac{1}{2} \psi^{(2)}(y+ m) + \frac{1}{3m} ( \psi^{(1)}(y) - \psi^{(1)}(y+ m) ) \nonumber \\
&+& \frac{1}{3 m^2} (\psi(y+ m) - \psi(y)) \biggr] \biggr], \nonumber \\
\Lambda_\perp &=& \frac{3}{4} \biggl[-\frac{1}{60} D_1(0) \psi^{(1)}(y) + \sum_{m \geq 1} D_1(m) \biggl[ \psi^{(1)}(y) - \psi^{(1)}(y+m) + \frac{1}{3m} ( \psi(y+m) - \psi(y)) \biggr] \biggr].
\end{eqnarray}

\begin{figure}[b]
\scalebox{0.6}[0.6]{\includegraphics{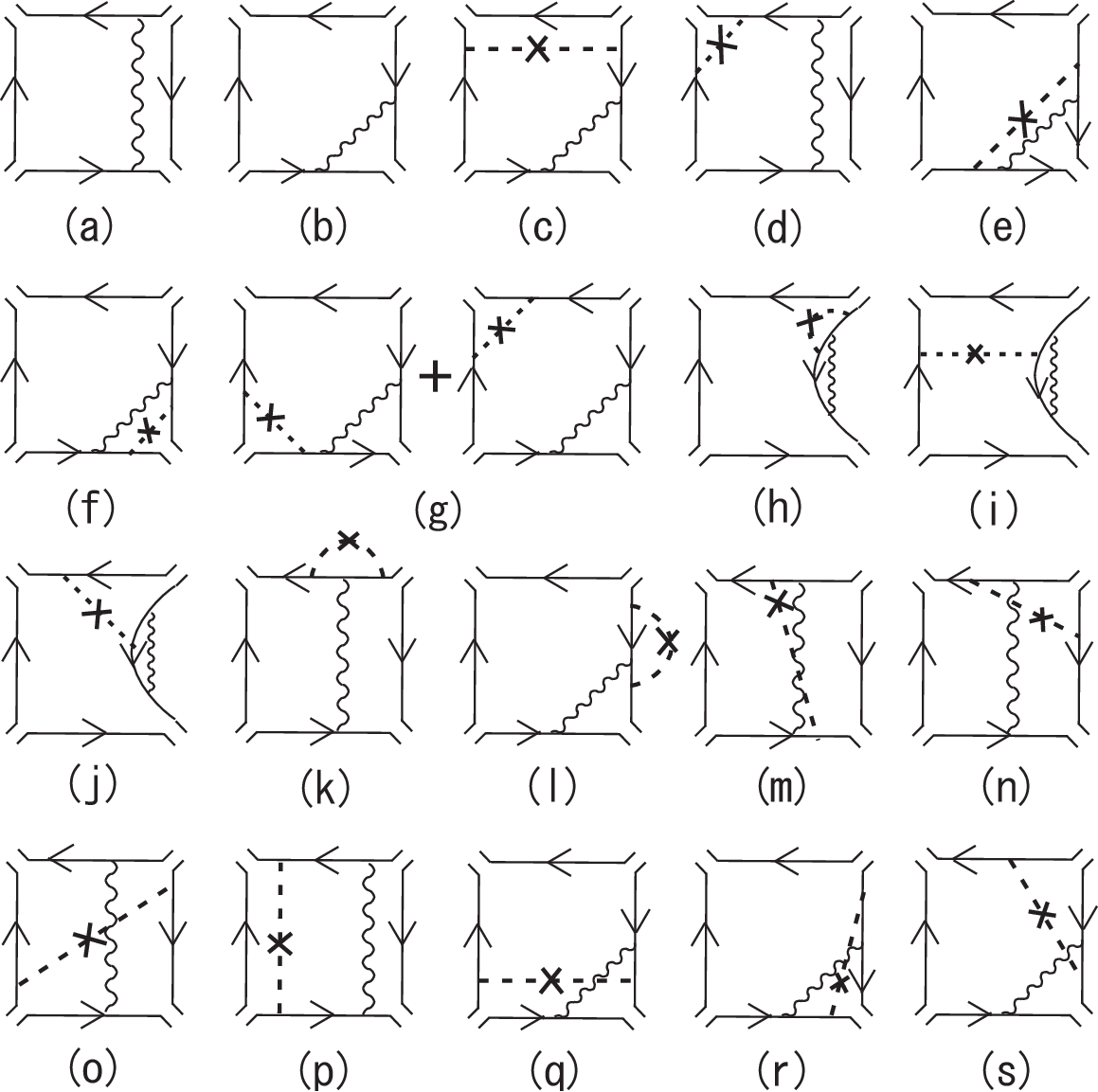}}
\caption{Diagrams expressing the SC contributions of O($\delta_u$) to the GL quartic terms. Here, regarding the figures (a) and (b), picking up just O($\delta_u$) terms from the self energy insertion there is implied, while the remaining ones express the terms with an impurity-induced vertex correction accompanied by the parameter $\delta_u$.}
\label{fig.6}
\end{figure}

Now, the anisotropic SC corrections to the $\beta_j$ parameters will be explained. According to the explanation in sec.III, we have contributions to $\beta_j$ from both $F_s^{(2)}$ and $F_s^{(4)}$ even in this anisotropic case. The latter, which is more divergent in the limit ${\overline I} \to 1$ than the terms accompanied by $D_1$ in the former, has been examined by substituting eqs.(\ref{deltachi2}) and (\ref{deltachiani2}) into eq.(\ref{quarfreesf}) and is found to be one or two orders of magnitude smaller than the contributions from $F_s^{(2)}$ for the values ${\overline I} \leq 0.9$ appropriate for liquid $^3$He below 30 (bar). Based on this fact consistent with the results seen in Ref.\cite{Aoyama}, we focus hereafter on the diagrams contributing to $F_s^{(2)}$ whic are described in Fig.6. 

As in the case of examining Fig.5, we have dropped diagrams to be absorbed into the weak-coupling terms in describing Fig.6. We wil not explain derivation of all SC diagrams in Fig.6, and rather we focus on diagrams with dominant contributions to the $\beta_j$ parameters and simply describe their expressions. Like in the impurity-scattering contributions to $\beta_j$ in the isotropic case and in the anisotropy effects on the GL quadratic term, the dominant anisotropic SC contributions to $\beta_j$ seem to arise from the diagrams (h), (i), and (j), which are based on a self energy terms in clean limit and show a crossing between an impurity line carrying $\delta_u$ and a paramagnon propagator. As in eqs.(\ref{isochi(1)var}) and (\ref{isochi(23)}), these diagrams are also enhanced as a result of their dependences on the high energy cutoff. The contributions to $\langle \delta \chi_{\mu,\nu}({\bf q},\omega) \rangle_{\hat q}$ of these diagrams are expressed 
as 
\begin{widetext}
\begin{eqnarray}
4 &\times& {\rm Fig.6 (h)} = - T \delta_{\mu,\nu} \sum_\varepsilon \int_{\bf p} \int_{\bf p'} \frac{\delta_u}{2 \pi \tau_o N(0)} ({\hat p}_z^2 + ({\hat p}'_z)^2) (|{\bf d}({\bf p})|^2)^2 \langle {\cal G}_{{\bf p}'+{\bf q}}(\varepsilon+\omega) {\cal G}_{{\bf p}'}(\varepsilon) {\cal G}_{{\bf p}+{\bf q}}(\varepsilon+\omega) {\cal G}_{\bf p}^3(\varepsilon) {\cal G}_{-{\bf p}}^2(-\varepsilon) \rangle_{\hat q} \nonumber \\
&=& - 12 \pi^2 \delta_{\mu,\nu} \frac{\delta_u N(0) T}{2 \pi \tau_o} \sum_\varepsilon \theta(\varepsilon(\omega-\varepsilon)) \frac{1}{(2 \varepsilon)^4} \biggl\langle \biggl\langle \biggl\langle ({\hat p}_z^2 + ({\hat p}'_z)^2) \frac{{\hat p}_i {\hat p}_j {\hat p}_k {\hat p}_l \omega^2}{[({\bf v}_{\bf p}\cdot{\bf q})^2 + \omega^2][({\bf v}_{{\bf p}'}\cdot{\bf q})^2 + \omega^2]} \biggr\rangle_{\hat p} \biggr\rangle_{\hat p'} \biggr\rangle_{\hat q} A_{\mu,i}^{\ast}A_{\nu,k}^{\ast}A_{\mu,j}A_{\nu,l} 
\nonumber \\
&=& -\frac{\pi^4}{240} \delta_{\mu,\nu} \frac{T N(0) \delta_u}{2 \pi \tau_o v_{\rm F}^2 q^2} \sum_{\varepsilon > 0} \biggl(\frac{1}{\varepsilon^4} - \frac{1}{(\varepsilon+|\omega|)^4} \biggr) [ (A_{\mu,i}^* A_{\mu,i})^2 + A^*_{\mu,i} A^*_{\nu,i} A_{\mu,j} A_{\nu,j} + A^*_{\mu,i} A_{\nu,i} A^*_{\nu,j} A_{\mu,j} \nonumber \\
&+& (A_{\mu,i}^* A_{\mu,i} A_{\nu,z}^* A_{\nu,z} + A^*_{\mu,i} A^*_{\nu,i} A_{\mu,z} A_{\nu,z} + A^*_{\mu,i} A_{\nu,i} A^*_{\nu,z} A_{\mu,z} + {\rm c.c.}) 
], \nonumber \\
2 &\times& {\rm Fig.6 (i)} = - T \delta_{\mu,\nu} \sum_\varepsilon \int_{\bf p} \int_{{\bf p}'} \frac{\delta_u}{4 \pi \tau_o N(0)} ({\hat p}_z^2 + ({\hat p}'_z)^2) [(|{\bf d}({\bf p})|^2) \, \langle {\cal G}_{{\bf p}+{\bf q}}(\varepsilon+\omega) {\cal G}_{\bf p}^2(\varepsilon) {\cal G}_{-{\bf p}}(-\varepsilon) ] \nonumber \\
&\times& [(|{\bf d}({\bf p}')|^2) {\cal G}_{{\bf p}'+{\bf q}}(\varepsilon+\omega) {\cal G}_{{\bf p}'}^2(\varepsilon) {\cal G}_{-{\bf p}'}(-\varepsilon) ] \rangle_{\hat q} \nonumber \\
&=& - \frac{\pi^4}{1120} \frac{T N(0) \delta_u}{2 \pi \tau_o v_{\rm F}^2 q^2} \sum_{\varepsilon > 0} \frac{1}{\varepsilon^4} \biggl[ - \frac{1}{6} (A_{\mu,i}^* A_{\mu,i})^2 - \frac{1}{6} A^*_{\mu,i} A^*_{\nu,i} A_{\mu,j} A_{\nu,j} + A^*_{\mu,i} A_{\nu,i} A^*_{\nu,j} A_{\mu,j} \nonumber \\
&+& \biggl( \frac{5}{6} A_{\mu,i}^* A_{\mu,i} A_{\nu,z}^* A_{\nu,z} + \frac{5}{6} A^*_{\mu,i} A^*_{\nu,i} A_{\mu,z} A_{\nu,z} + 2 A^*_{\mu,i} A_{\nu,i} A^*_{\nu,z} A_{\mu,z} + {\rm c.c.} \biggr) 
\biggr], \nonumber \\
4 &\times& {\rm Fig.6 (j)} = T \delta_{\mu,\nu} \sum_\varepsilon \int_{\bf p} \int_{{\bf p}'} \frac{\delta_u}{\pi \tau_o N(0)} {\hat p}_z {\hat p}'_z d_\mu^*({\bf p}') d_\mu({\bf p}) (|{\bf d}({\bf p})|^2) \, \langle {\cal G}_{{\bf p}'+{\bf q}}(\varepsilon+\omega) {\cal G}_{{\bf p}'}(\varepsilon) \nonumber \\
&\times& {\cal G}_{-{\bf p}'}(-\varepsilon) {\cal G}_{{\bf p}+{\bf q}}(\varepsilon+\omega) {\cal G}_{\bf p}^2(\varepsilon) {\cal G}_{-{\bf p}}^2(-\varepsilon) \rangle_{\hat q} \nonumber \\
&=& -\frac{\pi^4}{840} \delta_{\mu,\nu} \frac{T N(0) \delta_u}{2 \pi \tau_o v_{\rm F}^2 q^2} \sum_{\varepsilon > 0} \frac{1}{\varepsilon^4} \biggl[ (A_{\mu,i}^* A_{\mu,i})^2 + A^*_{\mu,i} A^*_{\nu,i} A_{\mu,j} A_{\nu,j} + A^*_{\mu,i} A_{\nu,i} A^*_{\nu,j} A_{\mu,j} \nonumber \\ 
&+& \frac{11}{2}(A_{\mu,i}^* A_{\mu,i} A_{\nu,z}^* A_{\nu,z} + A^*_{\mu,i} A^*_{\nu,i} A_{\mu,z} A_{\nu,z} + A^*_{\mu,i} A_{\nu,i} A^*_{\nu,z} A_{\mu,z} + {\rm c.c.}) 
\biggr].
\label{diaghij} 
\end{eqnarray}
\end{widetext}
By gathering eq.(\ref{diaghij}) and the contributions of other diagrams in Fig.6, up to the lowest order in $\delta_u$, the anisotropic terms in the total SC corrections to $\beta_j$, defined by $\delta \beta_j^{({\rm sc})}(\delta_u) = \beta_j^{({\rm sc})}(\delta_u) - \beta_{j, {\rm se}}^{({\rm sc})}(0) - \beta_{j, {\rm vc}}^{({\rm sc})}(0)$, become 
\begin{widetext}
\begin{eqnarray}
\delta \beta_2^{({\rm sc})}(\delta_u) &=& \frac{\delta_u \pi^4 {\overline I}^2}{4 \pi \tau_o T} {\overline \beta}_0 \biggl[ \pi^{-2} \biggl(\frac{12}{7} \Sigma_a - \frac{4}{7}\Sigma_b - \frac{8}{3}\Sigma_c \biggr) + 3 \Lambda_h - \frac{8}{3}\Lambda_l + 4 \Lambda_k + \frac{9}{14} \biggl(\Lambda_i + \Lambda_m - \frac{2}{3} \Lambda_q \biggr) - \frac{3}{7} \Lambda_o \nonumber \\
&+& \frac{6}{7} \biggl(\Lambda_j + 2 \Lambda_n - \frac{1}{6} (\Lambda_r+\Lambda_s) \biggr) + \frac{64}{3 \pi^2} \Lambda_c + \frac{4}{7 \pi^2} (3 \Lambda_b + 5 \Lambda_p) \biggr], \nonumber \\
\delta \beta_3^{({\rm sc})}(\delta_u) &=& \delta \beta_2^{({\rm sc})}(\delta_u) + \frac{\delta_u \pi^4 {\overline I}^2}{4 \pi \tau_o T} {\overline \beta}_0 \biggl[ \frac{8}{3 \pi^2} \Sigma_c - \frac{3}{4}(\Lambda_i + \Lambda_m) + \frac{1}{2} \Lambda_q - \frac{64}{3 \pi^2} \Lambda_c  \biggr], \nonumber \\
\delta \beta_4^{({\rm sc})}(\delta_u) &=& \delta \beta_3^{({\rm sc})}(\delta_u) + 3 \frac{\delta_u \pi^4 {\overline I}^2}{4 \pi \tau_o T} {\overline \beta}_0 \Lambda_o, \nonumber \\
\delta \beta_{2z}^{({\rm sc})}(\delta_u) &=& \frac{\delta_u \pi^4 {\overline I}^2}{4 \pi \tau_o T} {\overline \beta}_0 \biggl[ \pi^{-2} \biggl(\frac{24}{7}\Sigma_a - \frac{8}{7}\Sigma_b- \frac{8}{3} \Sigma_c + 16 \Sigma_d - \frac{4}{3} \biggl(\Sigma_e + \frac{1}{2}\Sigma_f + \frac{3}{2} \Sigma_g \biggr) \biggr) + 3\Lambda_h - \frac{8}{3}\Lambda_l + 4 \Lambda_k + \frac{9}{7}(\Lambda_i+\Lambda_m) \nonumber \\
&-& \frac{6}{7}\Lambda_q + \frac{33}{7}(\Lambda_j + 2\Lambda_n) - \frac{11}{14}(\Lambda_r+\Lambda_s) + \frac{15}{7}\Lambda_o+\frac{24}{7 \pi^2} \Lambda_b + \frac{64}{3 \pi^2}\Lambda_c + \frac{12}{7 \pi^2} \Lambda_p + \frac{16}{3 \pi^2}(2\Lambda_e+\Lambda_f+3\Lambda_g)  \biggr], \nonumber \\
\delta \beta_{3z}^{({\rm sc})}(\delta_u) &=& \delta \beta_{2z}^{({\rm sc})}(\delta_u) + \delta \beta_3^{({\rm sc})}(\delta_u) - \delta \beta_2^{({\rm sc})}(\delta_u), \nonumber \\
\delta \beta_{4z}^{({\rm sc})}(\delta_u) &=& \delta \beta_{3z}^{({\rm sc})}(\delta_u) + \delta \beta_4^{({\rm sc})}(\delta_u) - \delta \beta_3^{({\rm sc})}(\delta_u), 
\label{delbetaani}
\end{eqnarray}
\end{widetext}
and $\delta \beta_1^{({\rm sc})}(\delta_u) = \delta \beta_5^{({\rm sc})}(\delta_u) = \delta \beta_{1z}^{({\rm sc})}(\delta_u) = \delta \beta_{5z}^{({\rm sc})}(\delta_u) = 0$. 
Here, $\delta \beta_{jz}^{({\rm sc})}$ is the SC correction to $\beta^{({\rm wc})}_{jz}$ introduced in sec.II, and the contribution from each diagram in Fig.6 is expressed in terms of the index specifying each diagram in the coefficients $\Sigma$ and $\Lambda$ of which the detailed expressions are shown in Appendix. 

\section{Results}

In this section, pressure v.s. temperature ($P$-$T$) phase diagrams describing possible superfluid phases of liquid $^3$He in a globally anisotropic aerogel are examined using the anisotropic SC corrections obtained in $\S$ 3. 
The free energy is calculated by gathering the expressions, (\ref{eq:quadr}), (\ref{eq:quatrfr}), (\ref{betawcbulk}), (\ref{bulkbetasc}), (\ref{isobetasc}), (\ref{aniquadrsc}), (\ref{aniquadrvc}), and (\ref{delbetaani}). The pairing states examined in the present work as a possible superfluid state in an uniaxially and globally anisotropic aerogel are the following five pairing states; polar, planar, A$_{\rm Z}$ with ${\hat {\bf l}}$-vector parallel to the anisotropy axis ${\hat z}$, anisotropic A$_{\rm XY}$ with ${\hat {\bf l}} \perp {\hat z}$, the biaxial ESP \cite{Sauls,yang}, and the anisotropic B states, all of which are in the category of the unitary states \cite{VW}. The planar phase never becomes the state with the lowest energy in any situation we have studied and will not appear hereafter in describing phase diagrams. Except the biaxial ESP state, the remaining states have been examined in the previous study \cite{AI06} where no anisotropic SC effects have been considered. The biaxial ESP state will be defined later in the text. 

Throughout our analysis, the known experimental data on the pressure dependences of $E_{\rm F}$ and $T_{c0}$ in the bulk liquid will be used. On the other hand, we have three microscopic parameters other than the anisotropy parameter $\delta_u$ in the present approach based on the paramagnon model of the SC corrections and the Born approximation of the impurity scattering; the averaged mean free path of quasiparticles $l_{\rm mf} = v_{\rm F} \tau_o$, the dimensionless interaction parameter ${\overline I} \equiv I N(0)$, and a high energy cutoff $E_c$ for the normal paramagnon propagator. First, weakly $P$-dependent ${\overline I}$-values have been assumed to study phase diagrams based on known ${\overline I}$-values appropriate for obtaining the phase diagram of the bulk liquid (see the captions of Fig.7 and 8). We have verified that such ${\overline I}$-values do not lead to emergence of the A-phase at least below 30 (bar) in the isotropic case with $\delta_u=0$. Further, the $\xi_0/l_{mf} = (2 \pi \tau_o T_{c0})^{-1}$-values have been chosen by following the previous works \cite{AI06,Aoyama}. In the figures in this section, $\xi_0/l_{mf} = 0.35$ was commonly used. In contrast, we have no knowledge on appropriate values of $E_c$, and $E_c$ inevitably becomes one of a free parameter in the present approach. Below, we show results following from two values of $E_c$ which result in remarkably different phase diagrams from each other. The cutoff $E_c$ is introduced into the expressions by assuming the integer $m$ to be summed over the values between zero and $E_c/(k_{\rm B} T)$. 

In Fig.7, the resulting two phase diagrams, (a) and (b), obtained in terms of the fixed $E_c/k_{\rm B}=160$(mK) are shown. The phase diagram Fig.7(a) at a relatively low anisotropy $\delta_u=-0.15$ is similar to the previous ones \cite{yang} obtained with no anisotropic SC effects. One of the new features in Fig.7(a) is the temperature width of the polar phase region which clearly diminishes with increasing $P$ reflecting the anisotropic SC effect in eq.(\ref{aniquadrvc}). 
As will be shown later, the reduction of the polar phase region at higher $P$ implies the presence at a higher pressure of a critical point at which the polar (and other pairing states) meets with the A-phase with ${\bf l}$-vector parallel to the anisotropy axis ${\hat z}$, denoted as ${\rm A}_z$. Above this critical pressure, the polar phase would be absent, and the A$_z$-phase become the high temperature pairing state. 
Figure 7 (b) is the corresponding phase diagram in the case with a larger anisotropy. It is found that an increase of the stretched anisotropy pushes the polar phase region up to higher temperatures, so that $T_c$ is increased by the stretched anisotropy, and pushes the B phase region down to lower temperatures. In particular, it is a remarkable feature that the superfluid region at lower pressures is expanded in temperature by the stretched anisotropy : Although the $T_c$-curve in the isotropic limit (the dashed curve) in Fig.7 implies the presence of a quantum critical positive pressure below which the normal state at zero temperature is present, the polar pairing region growing with increasing the anisotropy pushes the quantum critical pressure down to the negative pressure side. Together with these features, the weak $P$-dependence of $T_c(P)$ at higher $P$-values in Fig.7 (b) seems to be qualitatively consistent with the data in Ref.\cite{DmitrievLT}. On the other hand, the feature that $dT_c/dP <0$ at high pressures is not compatible with the experimental phase diagrams \cite{DmitrievLT}. 

However, to compare results of the present theory with experimental ones, it will be reasonable to take account of a pressure dependence of the high energy cutoff $E_c$ for the normal spin-fluctuation. We will assume $E_c$ to be scaled like $E_{\rm F}\sqrt{1-{\overline I}}$ based on the expression of the denominator of the paramagnon propagator. Figure 7 (c) is one example of the phase diagrams in the case with a large anisotropy obtained in terms of the $P$-dependent $E_c$. Reflecting the fact that $D_2(|m|)$ is not sensitive to $E_c$, this change of $E_c$ does not affect much the low pressure region in the phase diagram. Therefore, the already-mentioned enhancement of the polar phase at lower $P$ due to the stretched anisotropy is one of definite results in the present theory and a tendency consistent with the observation \cite{DmitrievLT}, while it has not been found in the essentially weak-coupling theory \cite{yang}. On the other hand, the $T_c(P)$-portion with $dT_c/dP < 0$ at high pressures in Fig.7 (b) is changed to that with a positive slope by the $P$-dependence of $E_c$, so that details of $P$-dependence of the polar phase region seem to be inevitably obscured with no knowledge on a proper $E_c(P)$. 

\begin{figure}[t]
\scalebox{0.4}[0.4]{\includegraphics{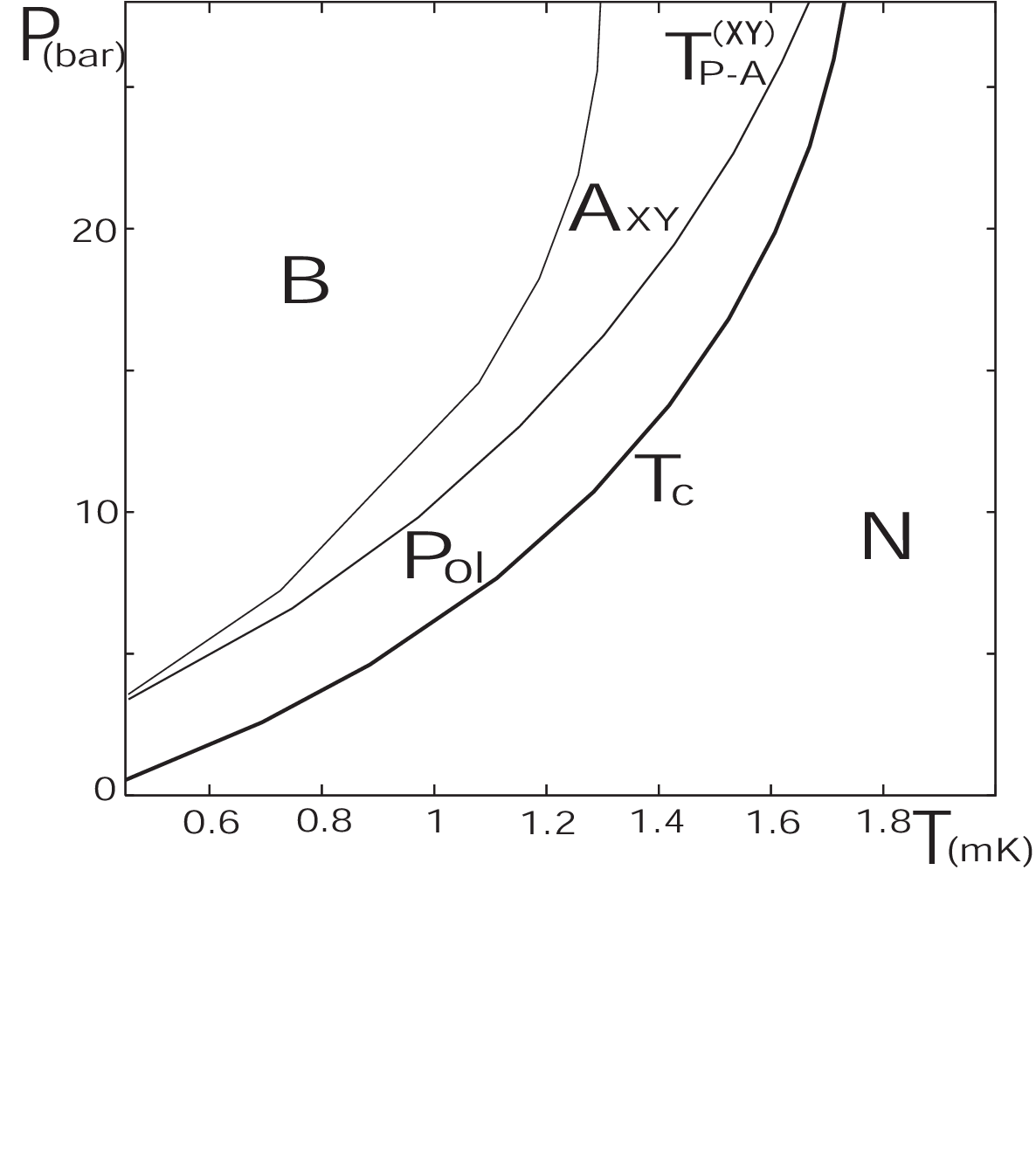}}
\scalebox{0.4}[0.4]{\includegraphics{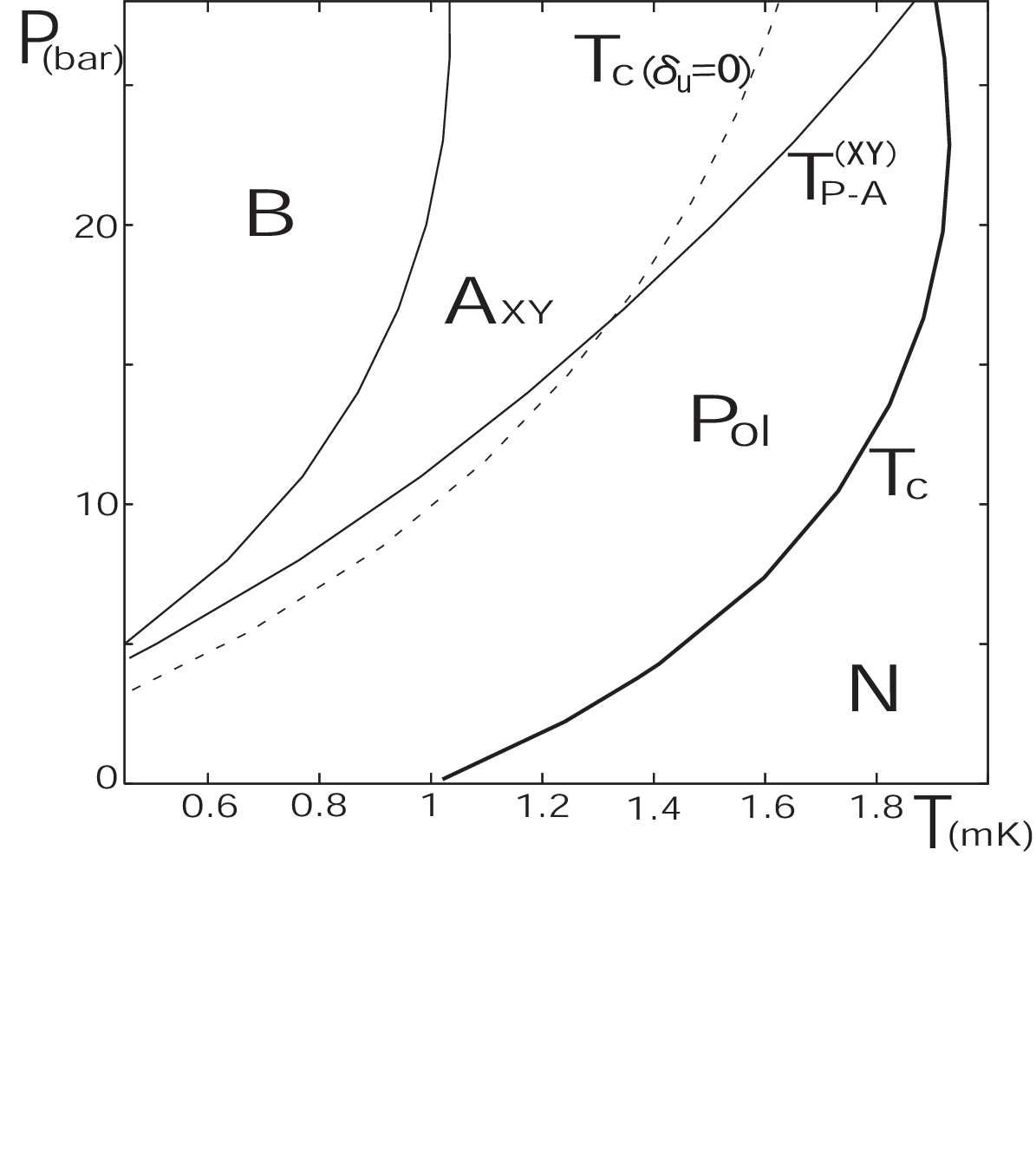}}
\scalebox{0.4}[0.4]{\includegraphics{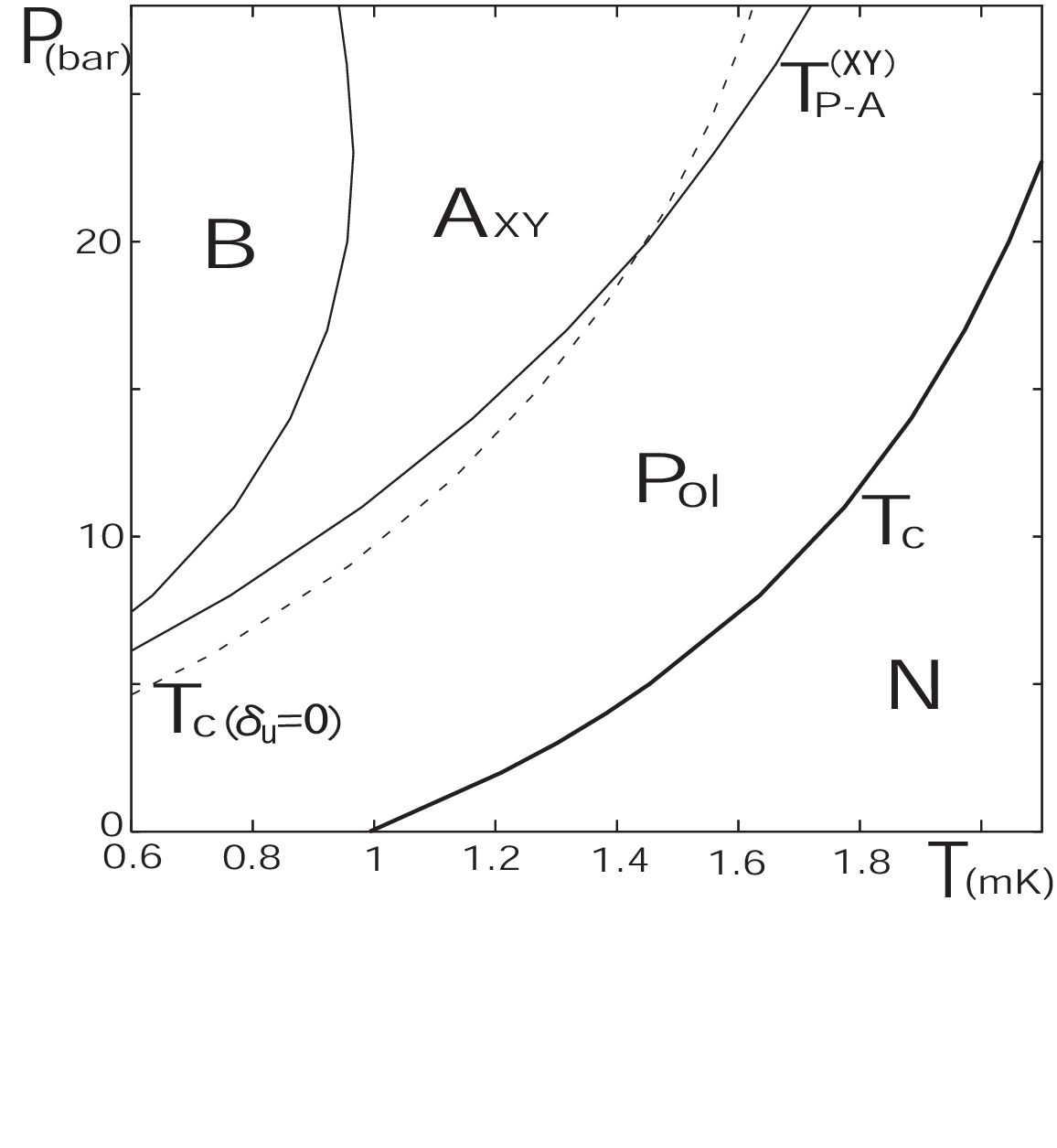}}
\caption{Resulting $P$-$T$ phase diagram in the stretched case ($\delta_u < 0$) obtained by use of ${\overline I}=0.5+0.005 P(bar)$. The fixed value $E_c/k_{\rm B} = 160$(mK) is used in (a) and (b), while we have used $E_c = 0.147 E_{\rm F}\sqrt{1 - {\overline I}}$ in (c). The anisotropy value $\delta_u$ is -0.15 in (a) and -0.7 in (b) and (c). }
\label{fig.7}
\end{figure}

Different types of phase diagrams in the uniaxially stretched aerogel ($\delta_u < 0$), which result from a larger $E_c(P)$, are shown in Fig.8. Due to the use of the larger $E_c$, the critical point on $T_c(P)$ at which the polar, A$_z$, and other pairing phases meet with one another is uncovered in the pressure range where experiments are usually performed. The obtained results suggest that, if the critical point on $T_c$ moves down to the $T=0$ limit, the resulting phase diagram would become consistent with that seen in Ref.\cite{HalperinNP} as follows: First, the superfluid phase realized upon cooling from the normal phase is the A$_z$ one with ${\hat {\bf l}}$-vector locked along the axis ${\hat z}$. 
In the present stretched case at higher pressures than the critical point, the low temperature superfluid phase is still the B phase at low enough anisotropy. For larger anisotropy values, however, another low temperature phase, called the biaxial ESP phase in Ref.\cite{Sauls}, can be realized. 

To explain the content on the possible biaxial ESP phase, the results in Refs.\cite{Sauls,yang} will be reviewed here: 
As far as no spatial inhomogenuity is present, the order parameter in the biaxial ESP state is written as $A_{\mu,j} = {\hat d}_\mu(a_1 {\hat x} + {\rm i} a_2 {\hat y} + p {\hat z})_j$. That is, this state can be regarded as an A-phase with its ${\hat {\bf l}}$-vector tilted in ${\hat z}$-${\hat x}$ plane from the anisotropy axis ${\hat z}$. Such a tilt of ${\hat {\bf l}}$-vector from ${\hat z}$ will not occur as far as the A$_{\rm XY}$ state does not become favorable over the A$_{\rm Z}$ one at lower 
temperatures. In fact, although this A$_{\rm Z}$ phase is also realized in uniaxially compressed aerogels at {\it lower} pressures \cite{yang} below the critical pressure,  the A$_{\rm XY}$ state is not stabilized there even at lower temperatures, and no ordering to the biaxial state is found. This picture strongly suggests that the appearance of the biaxial ESP state in a stretched aerogel at lower temperatures in Ref.\cite{HalperinNP} is a direct consequence of the sign reversal of the anisotropy primarily occuring in the GL-quadratic term at higher pressures. As shown elsewhere, the effective GL free energy on the ordering of a biaxial ESP state from the A$_{\rm Z}$ state takes the form \cite{yang}
\begin{equation}
f_{\rm eff}^{(bx)} = \biggl(\alpha_z - \frac{\beta_{245,z}}{\beta_{245}} \alpha \biggr) |p|^2 - \bigg(\frac{\beta_{245,z}^2}{\beta_{245}} + \frac{\beta_{13,z}^2}{\beta_{13}} \biggr) |p|^4, 
\label{effp}
\end{equation}
where $\alpha_z=\alpha_z^{({\rm wc})} + \alpha_z^{({\rm sc})}|_{\rm se} + \alpha_z^{({\rm sc})}|_{\rm vc}$, $\alpha = \alpha^{({\rm wc})} + \alpha^{({\rm sc})}|_{\rm se} + \alpha^{({\rm sc})}|_{\rm vc}$, $\beta_j=\beta_j^{({\rm wc})} + \beta_j^{({\rm sc})}$, $\beta_{jz}=\beta_{jz}^{({\rm wc})} + \beta_{jz}^{({\rm sc})}$, and $\beta_{ijk,z}=\beta_{iz}+\beta_{jz}+\beta_{kz}$. The biaxial ESP state is realized when $|p|^2 \propto ||a_1|^2 - |a_2|^2|$ is nonvanishing. It is clear that the quartic term of eq.(\ref{effp}) has a negative sign. It means that the transition between A$_{\rm Z}$ and the biaxial ESP states should be of first order and that, to make eq.(\ref{effp}) useful in calculation, the higher order ($|p|^m$ with $m \geq 6$) terms have to be incorporated. In the present work, just the left thin (red) solid curve $T_{\rm bx}^{(2)}(P)$ in each of Fig.8, at which the sign of the quadratic term of eq.(\ref{effp}) becomes negative on cooling, has been determined without going beyond the ordinary GL approach truncating to the quartic terms.  Then, we can only conclude here that the genuine biaxial ESP ordering from the A$_{\rm Z}$ state would occur at a {\it higher} temperature than the $T_{\rm bx}^{(2)}(P)$-line if the B phase does not have lower energy there. As Fig.8 (a) shows, the A$_{\rm Z}$-B transition curve $T^{({\rm Z})}_{\rm A-B}(P)$ which is determined by comparing free energies of the two phases with each other lies far above $T_{\rm bx}^{(2)}(P)$ in the case with a low anisotropy, suggesting that the B-phase should be realized at least in the intermediate temperature range in the case. Of course, in the limit of vanishing anisotropy, even the A$_{\rm Z}$ phase between the normal (N) and B phases is lost (see sec.III and also Fig.9 below). As in Fig.8 (b), however, the $T_{\rm bx}^{(2)}(P)$ curve approaches the $T^{({\rm Z})}_{\rm A-B}(P)$ one for a moderately large anisotropy and thus, it seems that a phase diagram with no B phase may be possibly realized. That is, to obtain a phase diagram qualitatively consistent with the observed one in Ref.\cite{HalperinNP}, not only a large energy transfer of the effective interaction (i.e., the paramagnon propagator) between the quasiparticles but also a large enough anisotropy are required.

\begin{figure}[t]
\scalebox{0.4}[0.4]{\includegraphics{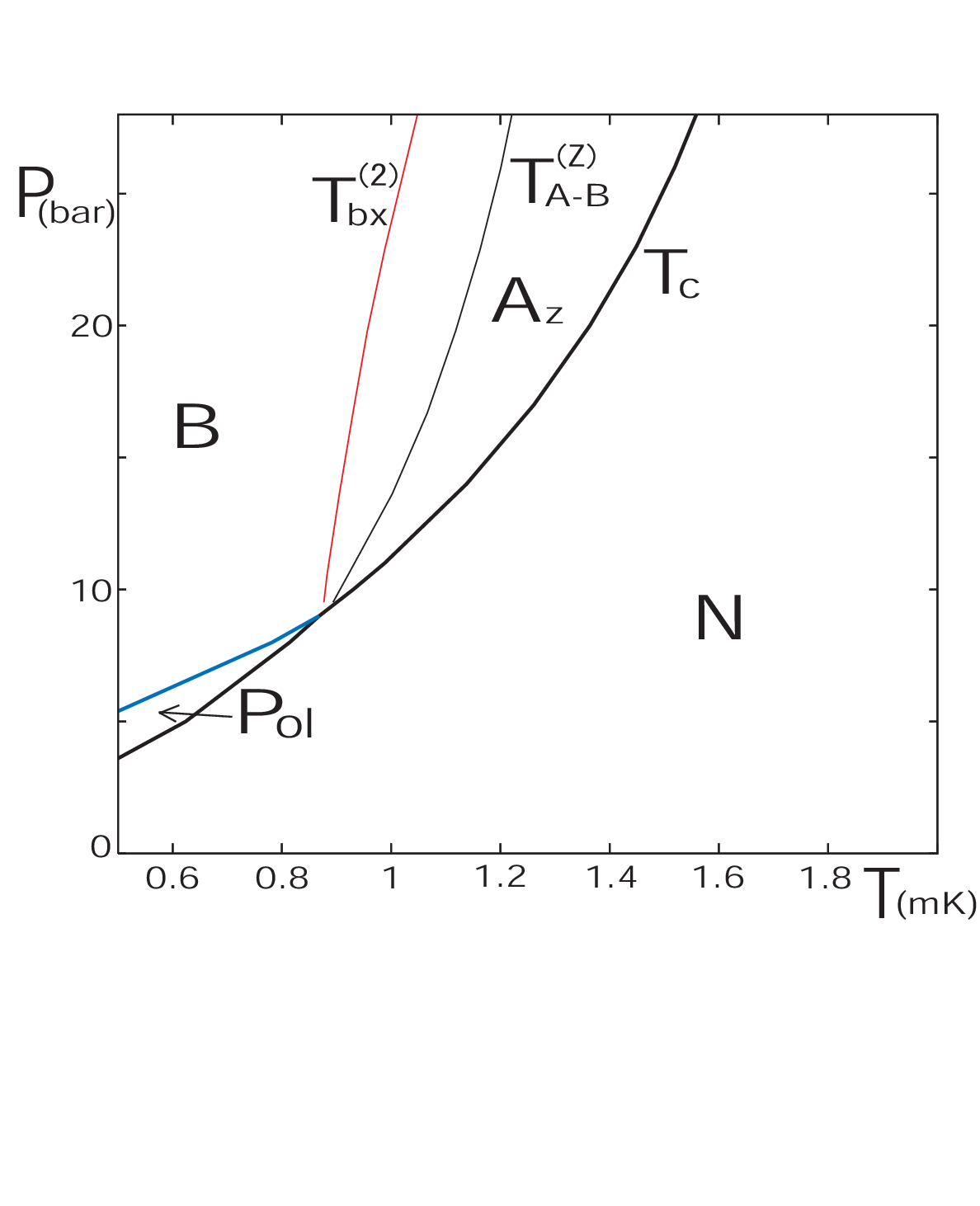}}
\scalebox{0.4}[0.4]{\includegraphics{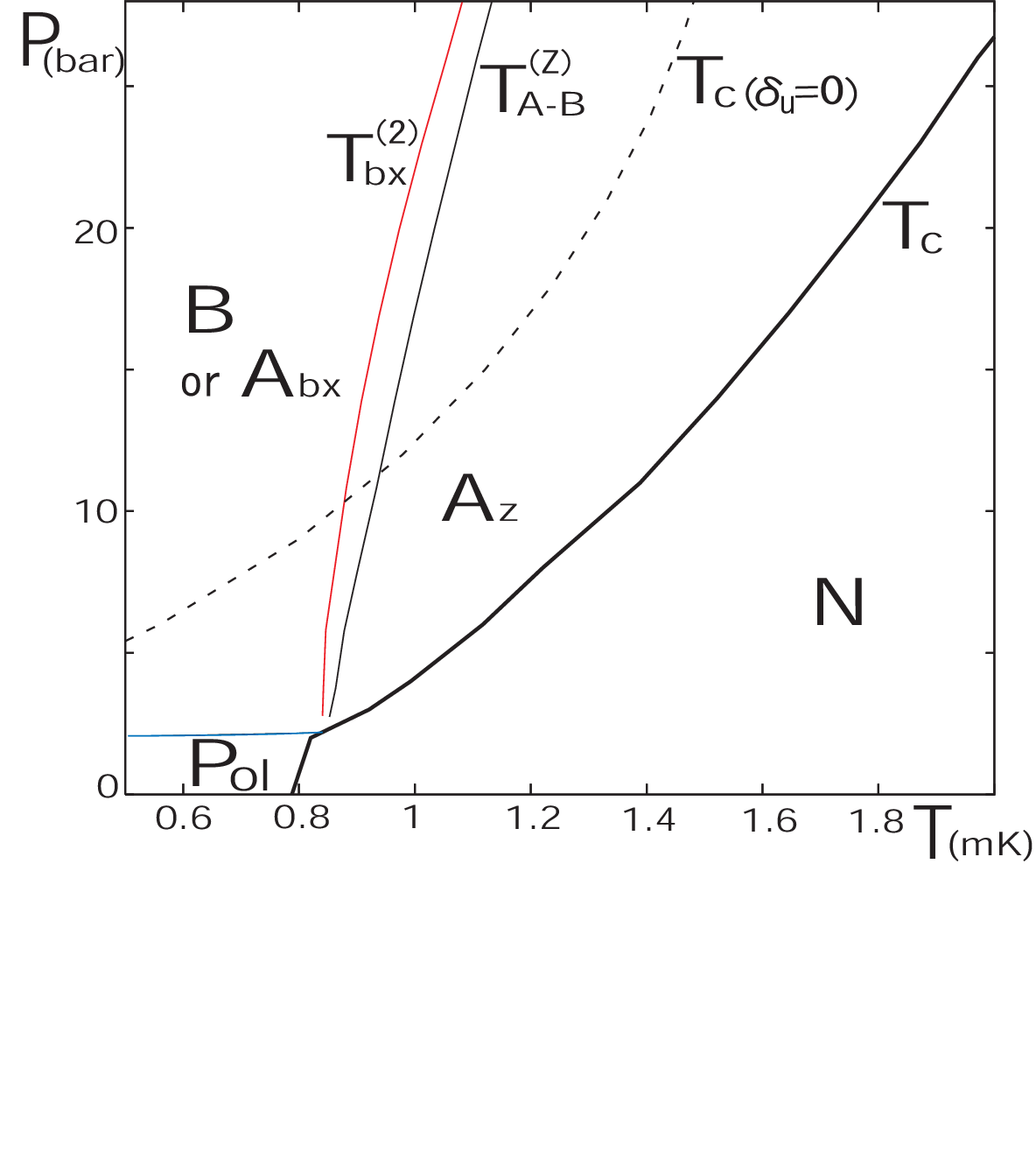}}
\caption{Resulting $P$-$T$ phase diagram in the uniaxially stretched case ($\delta_u < 0$) obtained by use of the parameter values ${\overline I}=0.63+0.005 P(bar)$ and $E_c = 0.87 E_{\rm F} \sqrt{1 - {\overline I}}$. The anisotropy value $\delta_u$ is -0.05 in (a) and -0.7 in (b). In the case (a) where the anisotropy is so weak that $T^{({\rm Z})}_{{\rm A}-{\rm B}}$ lies far above $T^{(2)}_{\rm bx}$, the biaxial ESP phase will not appear, ane the B phase is the only low temperature superfluid phase, while the B phase can be replaced by the biaxial ESP state for a larger stretched anisotropy.}
\label{fig.8}
\end{figure}

Finally, the superfluid phase diagrams following from the present approach in the case corresponding to the uniaxially compressed aerogel will be commented 
on. Even in this case, a critical pressure on $T_c(P)$ appears dependent on the high energy cutoff $E_c$. At lower pressures than the critical point, the system behaves as a 2D-like one. For a moderately large compressed anisotropy, the only possible superfluid phase becomes the A$_{\rm Z}$ state, and the B phase does not seem to appear on cooling. However, this situation does not correspond to that in a stretched aerogel in Ref.\cite{HalperinNP}, because, as already mentioned, the biaxial ESP state is never stabilized in this case. This situation below the critical pressure in a compressed aerogel has not been detected in experiments so far, and thus, we will not consider this low pressure region further 
here. 

In Fig.9, one example of the corresponding situation at higher pressures than the critical point is presented in the manner focusing on anisotropy dependences of possible phases. At this higher pressure side, the superfluid is 1D-like, and thus, within the present approach focusing on the unitary superfluid states, the polar pairing state is stabilized just below $T_c(P)$. In 1D-like case where an order along the axis is favored, the GL quadratic term makes the 1D form $A_{\mu,j} = d_\mu {\hat z}_j$ of the order parameter favorable near $T_c$. Further, because $\beta_{15}$ is negative with a large magnitude, energy is lowered when ${\hat d}_\mu$ is real rather than being complex as far as the additional $\beta_{jz}$-terms are ignored (see, e.g., eq.(\ref{eq:quatrfr})). That is, the polar pairing state which is one of ESPs should be realized in any 1D-like case at least for such a weak anisotropy that $\beta_{jz}$-terms are negligible even if a possibility of nonunitary states is taken into account. In Ref.\cite{HalperinPRL}, the presence of a non-ESP state just below $T_c$ has been suggested in a 1D-like situation of a uniaxially compressed aerogel in a vanishingly small magnetic field. Since any event close to $T_c$ will not be significantly affected by the GL-quartic terms, a drastic change of the GL-quadratic term would be necessary to justify such appearance of an unexpected phase near $T_c$. Figure 9 shows an extended polar phase at larger anisotropy, and it is unclear whether this tendency is changed by including an ingredient leading to a possibility of a nonunitary state.

\begin{figure}[t]
\scalebox{0.35}[0.35]{\includegraphics{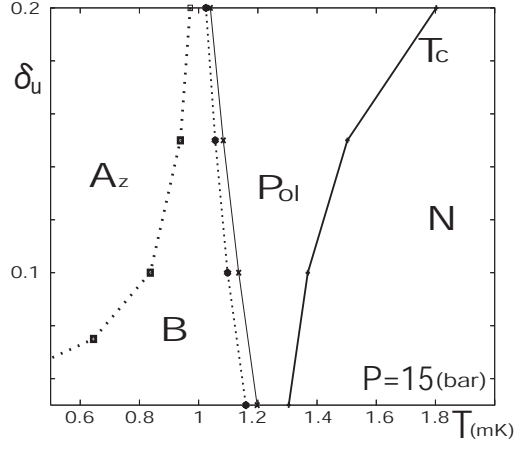}}
\caption{Anisotropy dependence of different transition temperatures in the case with uniaxially compressed aerogel with $\delta_u > 0$. There is quite a narrow temperature width of the $A_{\rm XY}$ phase between the polar and B phases. In the isotropic limit ($\delta_u \to 0$), the B phase becomes the only superfluid phase. The used energy cutoff is $E_c/k_{\rm B}=640$(mK).}
\label{fig.9}
\end{figure}

\section{Summary and Comments}

In the present work, the previous theory \cite{AI06} on superfluid phases of $^3$He induced by a globally anisotropic aerogel has been refined by incorporating anisotropy effects on the strong-coupling (SC) contributions to make calculated results comparable with recent experimental results \cite{DmitrievLT,HalperinNP} which have certainly detected novel superfluid phases in stretched or 1D-like aerogels. It is found that effects of the global anisotropy on the SC contributions are unexpectedly profound and can drastically change our understanding on emergent superfluid phases. First of all, the anisotropic SC effect tends to enhance the polar phase region as the pressure is lowered and, as seen in experiments in nematically-ordered aerogels \cite{DmitrievLT,Dmitriev}, significantly increases the superfluid transition temperature $T_c$ at lower pressures. Further, it has been found that high frequency contributions of the anisotropic SC effect to the free energy can change the sign of the uniaxial anisotropy at higher pressures. Consequently, the observed shrinkage of the polar phase at higher pressures in nematically-ordered aerogels \cite{DmitrievLT} and the 2D-like phase diagram in a 1D-like stretched aerogel \cite{HalperinNP} are qualitatively explained based on this sign reversal of anisotropy due to the SC effect. 

On the other hand, the present study for the superfluid phase diagram seen in an uniaxially compressed aerogel \cite{HalperinPRL} has not led to a convincing agreement with the experimental results. The presence of a {\it finite} threshold field for the emergence of the A$_{\rm XY}$ phase in Ref.\cite{HalperinPRL} suggests the presence of an additional non-ESP pairing state just below $T_c$ in the lowest fields which is more stable as the anisotropy is larger. The fact that this unexpected state, stable in the {\it apparently} 1D-like situation, is not the ESP-polar phase suggests that this novel state would be a nonunitary pairing sate which has not been considered in the present work based on the ordinary modelling identifying the aerogel structure with a nonmagnetic scatterer. However, in the experiments \cite{HalperinPRL} with no liquid $^4$He mixed \cite{comBill}, the layer of the solid $^3$He adsorbed on the local aerogel surface is active and might play a role of magnetic scattering centers, and hence, we might need to change our theoretical description on phase diagrams particularly close to $T_c$ where corrections to the mean field description work more effectively than deep in the ordered phase. Experimentally, it is hoped that the corresponding measurements in similar aerogels {\it with} liquid $^4$He mixed would be performed to verify whether the phase diagrams in \cite{HalperinPRL} and \cite{HalperinNP} are changed by the presence of active solid $^3$He layers or not. On the other hand, an extension of the present theoretical study to the case with the magnetic scatterings is left as a future work. 

\begin{acknowledgements}
This work was finantially supported by Grant-in-Aid for Scientific Research [No. 25400368] from MEXT, Japan. 
\end{acknowledgements}

\appendix

\section{Coefficients in $\delta \beta_{j}^{({\rm sc})}$}

The factors appearing in expressions (\ref{delbetaani}) of each diagram (a) - (s) in Fig.6 will be presented below. 
 
\begin{widetext}
\begin{eqnarray}
\Sigma_a &=& -\frac{D_2(0)}{24} \psi^{(4)}(y) - \sum_{m \geq 1} D_2(m) \biggl(\frac{1}{m^2} \psi^{(2)}(y) + \frac{2}{m^3}(\psi^{(1)}(y) - \psi^{(1)}(y+m)) \biggr), \nonumber \\
\Sigma_b &=& -\frac{D_2(0)}{6} \psi^{(4)}(y) - \sum_{m \geq 1} D_2(m) \biggl(\frac{1}{m}(\psi^{(3)}(y) - \psi^{(3)}(y+m)) + \frac{2}{m^2} \psi^{(2)}(y+m) + \frac{4}{m^3}(\psi^{(1)}(y) - \psi^{(1)}(y+m)) \biggr), \nonumber \\
\Sigma_c &=& \Sigma_e = \Sigma_g = -\frac{D_2(0)}{24} \psi^{(4)}(y) + \sum_{m \geq 1} D_2(m) \biggl(\frac{1}{3 m} (\psi^{(3)}(y) - \psi^{(3)}(y+m)) + \frac{1}{m^2} \psi^{(2)}(y+m) \nonumber \\
&+& \frac{2}{m^3}(\psi^{(1)}(y) - \psi^{(1)}(y+m)) + \frac{2}{m^4}(\psi(y+m) - \psi(y)) \biggr), \nonumber \\
\Sigma_d &=& -\frac{D_2(0)}{24} \psi^{(4)}(y) - \sum_{m \geq 1} D_2(m) \biggl(\frac{1}{m^2} \psi^{(2)}(y) + \frac{2}{m^3}(\psi^{(1)}(y) - \psi^{(1)}(y+m)) - \frac{2}{m^4}(\psi(y) - \psi(y+m)) \biggr), \nonumber \\
\Sigma_f &=& -\frac{D_2(0)}{24} \psi^{(4)}(y) + \sum_{m \geq 1} D_2(m) \biggl(-\frac{1}{m^2} \psi^{(2)}(y) + \frac{2}{m^3}(\psi^{(1)}(y) - \psi^{(1)}(y+m)) + \frac{6}{m^4}(\psi(y+m) - \psi(y)) \biggr), \nonumber \\
\Lambda_b &=& 2 D_1(0) \psi^{(3)}(y) + 2 \sum_{m \geq 1} D_1(m) \biggl( \psi^{(3)}(y) - \frac{2}{m} (\psi^{(2)}(y) - \psi^{(2)}(y+m)) + \frac{2}{m^2}(\psi^{(1)}(y) - \psi^{(1)}(y+m)) \biggr), \nonumber \\
\Lambda_c &=& \Lambda_e = \frac{2}{3} D_1(0) \psi^{(3)}(y) + \sum_{m \geq 1} D_1(m) \biggl(\frac{2}{3} \psi^{(3)}(y) - \frac{2}{m} (\psi^{(2)}(y) - \psi^{(2)}(y+m)) - \frac{2}{m^2} (\psi^{(1)}(y) +2 \psi^{(1)}(y+m)) \nonumber \\
&+& \frac{8}{m^3}(\psi(y+m) - \psi(y)) \biggr), \nonumber \\
\Lambda_f &=& \frac{D_1(0)}{3} \psi^{(3)}(y) + \sum_{m \geq 1} D_1(m) \biggl(- \frac{2}{m} (\psi^{(2)}(y) - \psi^{(2)}(y+m)) - \frac{2}{m^2} (\psi^{(1)}(y) +2 \psi^{(1)}(y+m)) + \frac{8}{m^3}(\psi(y+m) - \psi(y)) \biggr), \nonumber \\
\Lambda_g &=& \Lambda_i = \Lambda_j = \frac{D_1(0)}{6} \psi^{(3)}(y) + \sum_{m \geq 1} D_1(m) \frac{1}{3} \psi^{(3)}(y), \nonumber \\
\Lambda_h &=& \frac{1}{3}\sum_{m \geq 1} D_1(m) (\psi^{(3)}(y) - \psi^{(3)}(y+m)), \nonumber \\
\Lambda_k &=& 2 \sum_{m \geq 1} D_1(m) \frac{1}{m^2} \biggl( \psi^{(1)}(y) - \psi^{(1)}(y+m) + \frac{2}{m} (\psi(y+m) - \psi(y)) \biggr), \nonumber \\
\Lambda_l &=& \Lambda_k - \sum_{m \geq 1} D_1(m) \frac{1}{m}(\psi^{(2)}(y) - \psi^{(2)}(y+m)), \nonumber \\
\Lambda_m &=& \Lambda_n = \Lambda_o = \frac{D_1(0)}{6} \psi^{(3)}(y) + 4 \psi^{(1)}(y) \sum_{m \geq 1} D_1(m) \frac{1}{m^2}, \nonumber \\
\Lambda_p &=& \frac{D_1(0)}{6} \psi^{(3)}(y) + 2 \sum_{m \geq 1} D_1(m) \frac{1}{m^2} \biggl( 2 \psi^{(1)}(y+m) - \frac{3}{m} (\psi(y+m) - \psi(y)) \biggr), \nonumber \\
\Lambda_q &=& \Lambda_r = \Lambda_s = \frac{D_1(0)}{3} \psi^{(3)}(y) - 2 \psi^{(1)}(y) \sum_{m \geq 1} D_1(m) \frac{1}{m^2}. 
\end{eqnarray}
\end{widetext}


\begin{thebibliography}{99}
\bibitem{correc} The overall sign in eq.(3.3) in Ref.\cite{BSA} has to be corrected. This change does not affect other results there.
\end{thebibliography}
\end{document}